\documentclass[useAMS,usenatbib,usegraphicx]{mn2e}

\newcommand{\pin}{\sc pinocchio}
\newcommand{\gal}{\sc morgana}
\newcommand{\be}{\begin{equation}}
\newcommand{\ee}{\end{equation}}
\newcommand{\bea}{\begin{eqnarray}}
\newcommand{\eea}{\end{eqnarray}}
\newcommand{\msunyr}{M$_\odot$ yr$^{-1}$}
\newcommand{\msun}{M$_\odot$}
\newcommand{\kms}{km s$^{-1}$}
\newcommand{\surf}{M$_\odot$ pc$^{-2}$}

\title[The Effect of Feedback on the AGN population]
{The Effect of Stellar Feedback and Quasar Winds on the AGN population}

\author[Fontanot et al.]{Fabio Fontanot$^{1,2}$, Pierluigi Monaco$^{1,3}$ Stefano Cristiani$^3$ \& Paolo Tozzi$^3$\\ 
$^1$Dipartimento di Astronomia, Universit\`a di Trieste, via Tiepolo 11, 34131 Trieste, Italy \\
$^2$Max Planck Institute for Astronomy, K\"onigstuhl 17, D-69117 Heidelberg, Germany\\
$^3$INAF-Osservatorio Astronomico di Trieste, via Tiepolo 11, 34131 Trieste, Italy \\ 
email: fontanot@mpia-hd.mpg.de; monaco, cristiani, tozzi@oats.inaf.it}

\begin{document}
\date{Accepted ... Received ...}

\pagerange{\pageref{firstpage}--\pageref{lastpage}} \pubyear{}

\maketitle
\label{firstpage}

\begin{abstract}
In order to constrain the physical processes that regulate and
downsize the Active Galactic Nucleus (AGN) population, the predictions
of the MOdel for the Rise of GAlaxies aNd Active nuclei ({\gal}) are
compared to luminosity functions (LFs) of AGNs in the optical, soft
X-ray and hard X-ray bands, to the local BH--bulge mass relation, and
to the observed X-ray number counts and background.  We also give
predictions on the accretion rate of AGNs in units of the Eddington
rate and on the BH--bulge relation expected at high redshift.  We find
that it is possible to reproduce the downsizing of AGNs within the
hierarchical $\Lambda$CDM cosmogony and that the most likely
responsible for this downsizing is the stellar kinetic feedback that
arises in star-forming bulges as a consequence of the high level of
turbulence and leads to a massive removal of cold gas in small
elliptical galaxies.  At the same time, to obtain good fits to the
number of bright quasars we need to require that quasar-triggered
galactic winds self-limit the accretion onto BHs; however, the very
high degree of complexity of the physics of these winds, coupled with
our poor understanding of it, hampers more robust conclusions.  In all
cases, the predicted BH--bulge relation steepens considerably with
respect to the observed one at bulge masses $<10^{11}$ \msun; this
problem is related to a known excess in the predicted number of small
bulges, common to most similar models, so that the reproduction of the
correct number of faint AGNs is done at the cost of underestimating
their BH masses.  This highlights an insufficient downsizing of
elliptical galaxies, and hints for another feedback mechanism able to
act on the compact discs that form and soon merge at high redshift.
The results of this paper reinforce the need for direct investigations
of the feedback mechanisms in active galaxies, that will be possible
with the next generation of astronomical telescopes from sub-mm to
X-rays.

\end{abstract}



\section{Introduction}
\label{section:introduction}

In the last years evidence has grown of the importance of the
interaction between AGNs and their host galaxies.  This interaction is
now deemed to be of fundamental relevance for understanding the
assembly not only of the supermassive Black Holes (BH) responsible for
AGN activity, but also of the galaxies, in particular their spheroidal
components (ellipticals or spiral bulges).  Observations of nearby
galaxies show the existence of a well-defined correlation between the
mass or central velocity dispersion of bulges and the mass of the
hosted BHs (Kormendy \& Richstone, 1995; Magorrian et al., 1998; more
recent determinations are found, e.g., in Marconi \& Hunt, 2003 and
H\"aring \& Rix, 2004).  Moreover the observed mass function of these
BHs is consistent with that inferred from quasar\footnote{In this
paper we refer to luminous AGNs as quasars, with no reference to their
radio emission.}  luminosities, under simple assumptions for the
radiative efficiency and accretion rate (Soltan 1982; Cavaliere \&
Padovani 1988; Salucci et al. 1999; Yu \& Tremaine 2002; Shankar et
al. 2004; Marconi et al. 2004; see also Haiman, Ciotti \& Ostriker
2004). On the other hand, observations at high redshift highlight that
quasars and radio-loud AGNs are hosted in elliptical galaxies (see
e.g. Dunlop et al. 2003). 

An indirect evidence of the BH -- bulge connection relies in the
parallel evidence of downsizing of both populations.  The evolution of
the LFs of AGNs in the soft and hard X-ray bands reveals that the
number density of fainter objects peaks at a lower redshift with
respect to brighter ones (see i.e. Ueda et al. 2003; Hasinger, Miyaji
\& Schmidt 2005; Barger et al. 2005; La Franca et al. 2005).  This
evidence of downsizing is often referred to as the
``anti-hierarchical'' behavior of BH growth, and is confirmed by the
analysis of Merloni (2004), based on the fundamental plane of
accreting BHs (Merloni, Heinz \& di Matteo, 2003), and by Marconi et
al. (2004).  Recently, the GOODS collaboration determined the LF of
low-luminosity quasars at $z\sim 4$ (Cristiani et al. 2004; Fontanot
et al. 2006a), revealing a dearth of objects with respect to naive
extrapolations of Pure Luminosity Evolution models anchored to the
SDSS bright quasars (Fan et al. 2003). This evidence, confirmed by the
COMBO17 survey at higher luminosities (Wolf et al. 2003) is again in
line with the downsizing trend mentioned above.

A similar trend has longly been claimed by some authors to be observed
in galaxies (see, e.g., Cowie et al. 1996; Treu et al. 2005; Papovich
et al. 2006; Fontana et al. 2006), especially in ellipticals, whose
stellar mean ages, metal enrichment and level of alpha-enhancement are
known to correlate positively with the stellar mass (see, e.g. Thomas
1999; Renzini 2004).  These evidences point to a formation scenario
characterized by a quick burst of star formation rapidly followed by a
strong wind which expels the residual Inter-Stellar Medium (ISM); this
wind should take place earlier in more massive galaxies (Matteucci
1994) and may be triggered by the quasar shining (Granato et
al. 2001).  In this framework it is also possible to explain the high
metallicity of quasar hosts (Matteucci \& Padovani 1993; Hamann \&
Ferland 1999; Romano et al. 2002; D'Odorico et al. 2005).

The influence of AGNs may not be limited to galaxies: jets from radio
galaxies are now one of the most promising candidates for quenching
cooling flows in galaxy clusters (McNamara et al. 2005; Voit \&
Donahue 2005; Fabian et al. 2006; see also the simulations of Quilis,
Bower, \& Balogh 2001; Dalla Vecchia et al. 2004; Ruszkowski, Bruggen
\& Begelman 2004; Zanni et al. 2005; Sijacki \& Springel 2006), and
thus limit the mass of the most massive ellipticals (Benson et
al. 2003; Bower et al. 2006; Croton et al. 2006).

The problem of the demography of the AGN population in the context of
galaxy formation in $\Lambda$CDM models has been addressed by many
authors (see, e.g., Kauffmann \& Haehnelt 2000; Monaco, Salucci \&
Danese 2000; Cavaliere \& Vittorini 2000, 2002; Cattaneo 2001; Granato
et al. 2001, 2004; Mahmood, Devriendt \& Silk 2004; Bromley,
Somerville \& Fabian 2004; Menci et al. 2003, 2004, 2006; Cattaneo et
al. 2005; Vittorini et al. 2005; Bower et al. 2006; Croton et
al. 2006; Hopkins et al. 2006; Malbon et al. 2006).  In particular,
the downsizing of galaxies and AGNs is not straightforward to
reproduce in the hierarchical cosmogonic scenario, where larger dark
matter (DM) halos form on average later by the merging of less massive
halos.  As far as AGNs are concerned, Monaco, Salucci \& Danese (2000)
and later Granato et al. (2001; 2004) suggested that a delay in the
shining of quasars in small bulges, motivated by stellar feedback,
could explain this trend.  Unfortunately, a more precise assessment of
the success of the models listed above in predicting the downsizing of
AGNs is not easy in most cases, as only a few of these papers (like
Menci et al. 2004) explicitly compare to data in terms of the number
density of AGNs as a function of luminosity.

It is worth noting that in these papers different ideas have been
proposed on the origin of the BH -- bulge relation.  Several proposed
that the feedback from the AGN is able to self-limit the masses of
both spheroids and BHs, (see, e.g., Ciotti \& Ostriker 1997; Silk \&
Rees 1998; Haehnelt, Natarajan \& Rees 1998; Fabian 1999; Granato et
al. 2004; Murray, Quataert \& Thompson 2005; Monaco \& Fontanot 2005).
Such self-limitation is supported also by the results of N-body
simulations (Di Matteo et al. 2003, 2005; Kazantzidis et al. 2005;
Springel, Di Matteo \& Hernquist 2005).  However, the other authors
successfully reproduced the BH -- bulge relation simply assuming a
proportionality between star formation (in bulges) and accretion
rates, implicitly determined by the mechanism responsible for the
almost complete loss of angular momentum of the gas accreting on the
BH . The difference between the two approaches (self-limitation
or proportionality between star formation and accretion) was recently
clarified by Monaco \& Fontanot (2005).

This paper is the second of a series devoted to describe the {\gal}
code for the formation and evolution of galaxies and AGNs.  The code
is described in full detail in Monaco, Fontanot \& Taffoni (2006;
hereafter paper I), while the ability of the model to reproduce the
early formation and assembly of massive galaxies is demonstrated in
Fontana et al. (2006) and Fontanot et al. (2006b).
Here we focus on
the self-consistent modeling of accreting BHs and on their feedback on
the host galaxies.  By reproducing - or failing to reproduce - the
observed properties of the AGN population, in particular the LFs in
the optical, soft and hard X-ray bands, and the BH - bulge correlation
at $z=0$, we obtain valuable constraints on the physical processes
involved.  We also demonstrate our ability to roughly reproduce the
observed X-ray counts in the soft (0.5--2 keV) and hard (2--10 keV)
bands and background from 0.2 to 300 keV, and give predictions on the
BH -- bulge relation at high redshift and on the Eddington ratios of
the emitting objects.

The paper is organized as follows. In section~\ref{section:model} we
describe the main properties of the {\gal} model, the accretion onto
BHs and its feedback on the forming galaxy, the runs used in the
paper, and the procedure adopted to compute the AGN LFs, number counts
and X-ray background.  In section~\ref{section:results} we compare our
results with available data and give some predictions on AGN and BH
properties.  The results are discussed in
section~\ref{section:discussion}, while
section~\ref{section:conclusions} gives the conclusions.  Throughout
this work we assume the concordance $\Lambda$CDM cosmological model
with $\Omega_\Lambda=0.7$, $\Omega_m=0.3$, $H_0=70$ km s$^{-1}$
Mpc$^{-1}$, and $\sigma_8=0.9$.


\section{Modeling AGNs in \gal}
\label{section:model}

\subsection{\gal}
\label{section:morgana}

{\gal} has been presented preliminary by Monaco \& Fontanot
(2005), and is described in full length in paper I.  Here we give just
a brief outline of the main physical processes included
and discuss in some detail a few relevant points.

The {\gal} model follows the typical scheme of semi-analytic models,
with some important differences.  Each DM halo contains one galaxy for
each progenitor\footnote{
Each DM halo forms through the merging of many halos of smaller mass,
called progenitors. At each merging the largest halo survives (it
retains its identity), the others become substructure of the largest
one.  The main progenitor is the one that survives all the mergings.
The mass resolution of the box used for computing the merger trees
sets the smallest progenitor mass, as explained in
section~\ref{section:runs}.
}; the galaxy associated with the main progenitor is the central
galaxy.  Baryons in a DM halo are divided into three components,
namely a halo, a bulge and a disc. Each component contains three
phases, namely cold gas, hot gas and stars. For each component the
code follows the evolution of its mass, metal content, thermal energy
of the hot phase and kinetic energy of the cold phase.

The main processes included in the model are the following. 

(i) The merger trees of DM halos are obtained using the {\sc
pinocchio} tool (Monaco et al. 2002; Monaco, Theuns \& Taffoni 2002;
Taffoni, Monaco \& Theuns 2002).

(ii) After a merging of DM halos, dynamical friction, tidal stripping
and tidal shocks on the satellite (the smaller DM halo, with its
galaxy at the core) lead to a merger with the central galaxy or to
tidal destruction as described by Taffoni et al. (2003).

(iii) The evolution of the baryonic components is performed by
numerically integrating a system of equations for all the mass, energy
and metal flows.

(iv) The intergalactic medium infalling on a DM halo is shock-heated,
as well as the hot halo component of merging satellites (which is
given to the main halo) and that of the main halo in case of major
merger ($M_{\rm sat} > 0.2 \times M_{\rm tot}$). Following Wu, Fabian
\& Nulsen (2001), shock-heating is implemented by assigning to the
infalling gas a specific thermal energy equal to 1.2 times the
specific virial energy, $-0.5\, U_{\rm H}/M_{\rm H}$ (where $M_{\rm
H}$ and $U_{\rm H}$ are the mass and binding energy of the DM halo).

(v) The profile of the hot halo gas is computed at each time-step by
solving the equation for hydrostatic equilibrium with a polytropic
equation of state and an assumed polytropic index $\gamma_p=1.2$.

(vi) Cooling of the hot halo phase is computed by integrating the
contribution of each spherical shell, taking into account radiative
cooling and heating from (stellar and AGN) feedback from the central
galaxy: if $T_{g0}$ and $\rho_{g0}$ are the temperature and density of
the hot halo gas extrapolated to $r=0$, $\mu_{\rm hot}m_p$ its mean
molecular weight and $r_s=r_{\rm H}/c_{\rm nfw}$ the scale radius of
the halo (of radius $r_{\rm H}$ and concentration $c_{\rm nfw}$), then
the mass cooling flow $\dot{M}_{\rm co,H}$ results:

\be
\dot{M}_{\rm co,H} = \frac{4 \pi r_s^3\rho_{g0}}{t_{\rm cool,0}} 
\times {\cal I}(2/(\gamma_p-1))
\label{eq:coolingflow} 
\ee

\noindent where 

\be
t_{\rm cool,0} = \frac{3kT_g(r_{\rm cool}) \mu_{\rm hot} m_p}{2\rho_{g0}
(\Lambda_{\rm cool}-\Gamma_{\rm heat})}
\label{eq:cool_heat} 
\ee

\noindent
Here $\Lambda_{\rm cool}$ is the metal-dependent Sutherland \& Dopita
(1993) cooling function and the heating term $\Gamma_{\rm heat}$ is
computed assuming that the energy flow $\dot{E}_{\rm hw,H}$ fed back
from the galaxy (including SNe and AGN) is given to the cooling shell:

\be
\Gamma_{\rm heat} = \frac{\dot{E}_{\rm hw,H}}{4\pi r_s^3 
{\cal I}(2/(\gamma_p-1))} 
\left(\frac{\mu_{\rm hot}m_p}{\rho_{g0}}\right)^2
\label{eq:heating}
\ee

\noindent
Clearly these equations are valid if $\Gamma_{\rm heat}<\Lambda_{\rm
cool}$, otherwise no cooling flow is present.  In
equations~\ref{eq:coolingflow} and \ref{eq:heating} the integral
${\cal I}(\alpha)$ is defined as $\int_{r_{\rm cool}/r_s}^{c_{\rm
nfw}} \{ 1- a [1-\ln(1+t)/t ]\}^{\alpha}t^2 dt$, with $a=[3T_{\rm
vir}(\gamma_p-1)c_{\rm nfw}(1+c_{\rm nfw})]/ \{\gamma_p
T_{g0}[(1+c_{\rm nfw})\ln(1+c_{\rm nfw})-c_{\rm nfw}]\}$ ($T_{\rm
vir}$ being the virial temperature of the halo).  The cooling radius
$r_{\rm cool}$ is treated as a dynamical variable whose evolution
takes into account the hot gas injected by the central galaxy
($\dot{M}_{\rm hw,H}$):

\be
\dot{r}_{\rm cool} = \frac{\dot{M}_{\rm co,H}-\dot{M}_{\rm hw,H}}
{4\pi \rho_g(r_{\rm cool})r_{\rm cool}^2}\, .
\label{eq:drcool} \ee

(vii) The cooling gas flows into the cold halo gas phase; this is let
infall on the central galaxy on a dynamical time-scale (computed at
$r_{\rm cool}$).  This gas is divided between disc and bulge according
to the fraction of the disc that lies within the half-mass radius of
the bulge.  In case of a disc-less bulge the disc size is estimated as
$R_{\rm D}=0.7 \lambda r_{\rm H}$, where $\lambda$ is the spin
parameter of the DM halo; a more precise computation is performed as
explained in the next point when the disc accumulates a significant
amount of mass.

(viii) The gas infalling on the disc keeps its angular momentum; disc
sizes are computed with an extension of the Mo, Mao \& White (1998)
model that includes the contribution of the bulge to the disc rotation
curve.

(ix) Disc instabilities and major mergers of galaxies lead to the
formation of bulges.  We also take into account a possible disc
instability driven by feedback; this is explained in
section~\ref{section:stellarfb}.  In minor mergers the satellite mass
is given to the bulge component of the larger galaxy.

(x) Star formation and feedback in bulges and discs are inserted
following the model of Monaco (2004a).  For discs of gas surface density
$\Sigma_{\rm cold,D}$ and fraction of cold gas $f_{\rm cold,D}$ the
timescale for star formation $t_{\rm\star,D}$ is:

\be
t_{\rm\star,D} = 9.1\; \left(\frac{\Sigma_{\rm cold,D}}{1\ {\rm M}_\odot\ {\rm pc}^{-2}}\right)^{-0.73} 
\left(\frac{f_{\rm cold,D}}{0.1}\right)^{0.45}\ {\rm Gyr}
\label{eq:fbthin2}\ee

\noindent
Due to the correlation of $f_{\rm cold,D}$ and $\Sigma_{\rm cold,D}$
(galaxies with higher gas surface density consume more gas), this
relation is compatible with the Schmidt law.  For bulges the
straightforward Schmidt law is used:

\be
t_{\rm \star,B} = 4\;  \left(\frac{\Sigma_{\rm cold,B}}{1\ {\rm M}_\odot\ {\rm pc}^{-2}}\right)^{-0.4} \ {\rm Gyr}
\label{eq:schmidt} \ee

\noindent
In both cases, hot gas is ejected to the halo (in a hot galactic wind)
at a rate equal to the star-formation rate (as predicted by Monaco
2004a), though massive bulges with circular velocity $V_{\rm B}\ga300$
{\kms} are able to bind the $T\sim10^7$ K hot phase component.  The
thermal energy of this re-heated gas is not scaled to the DM halo
virial energy but is set equal to the energy of exploding SNe (assumed
to be $10^{51}$ erg) times an efficiency which is left as a free
parameter.  The best-fit value for the efficiency, 0.7, is very
similar to the 0.8 value suggested by Monaco (2004b), who estimated
the energy lost in the destruction of the host molecular cloud.

(xi) In star-forming bulges cold gas is ejected in a cold galactic
wind by kinetic feedback due to the predicted high level of turbulence
driven by SNe.  This is described in section~\ref{section:stellarfb}.

(xii) When the hot halo phase is heated beyond the virial temperature,
it can leave the DM halo in a galactic super-wind, at a rate:

\be
\dot{M}_{\rm hsw} = \left(1-\frac{f_{\rm wind} E_{\rm vir}}{E_{\rm hot,H}}\right) 
\frac{M_{\rm hot,H}}{t_{\rm sound}}  \label{eq:hotwind_m}
\ee

\noindent
where $E_{\rm hot,H}$ and $E_{\rm vir}$ are respectively the actual
and virial thermal energy of the hot phase ($E_{\rm vir}/M_{\rm
hot,H}=-0.5\, U_{\rm H}/M_{\rm H}$), $M_{\rm hot,H}$ its mass, $t_{\rm
sound}$ its sound-crossing time and $f_{\rm wind}$ is a free parameter
set to the value of 2.  A similar thing happens to the cold halo gas
when it is accelerated by stellar feedback.  To compute the time at
which the ejected gas falls back into a DM halo (and is shock-heated)
the merger history of the DM halo is scrolled forward in time until a
halo is met with circular velocity larger than the (sound or kinetic)
velocity of the gas at the ejection time.

(xiii) Metal enrichment is self-consistently modeled in the
instantaneous recycling approximation.

\subsection{Stellar feedback}
\label{section:stellarfb}

According to the model of Monaco (2004a), the regime of stellar
feedback in a galaxy depends mainly on the density and vertical
scale-length of the galactic system. In thin systems, like spiral
discs, the SNe exploding within a single molecular cloud give rise to
super-bubbles that quickly blow out of the system, so that only a
small fraction of their energy is injected into the ISM, while most
energy is given to the halo ($\sim80$ per cent according to Monaco
2004b).  This is called adiabatic blow-out regime, and results in an
ISM with thermal pressure of $\sim10^3$ K cm$^{-3}$ and a level of
turbulence quantified by a velocity dispersion of clouds of $\sim7$
\kms\ (see paper I).  In thick systems, characterized by a large
vertical scale-length or a high surface density, the super-bubbles do
not manage to blow out of the system, which is then in the so-called
adiabatic confinement regime, where most SN energy, both thermal and
kinetic, is injected into the ISM.  This results in higher thermal
pressure and velocity dispersion of clouds, well in excess of the
$\sim6$ \kms\ value found in spiral discs (see, e.g. Kennicutt 1989).

A reference set of parameters for stellar feedback is introduced
in paper I.  Most parameters are fixed by requiring to reproduce the
properties of local galaxies, or do influence the predictions of the
AGN population in a modest or rather predictable way, so we
concentrate on varying only two mostly relevant parameters, leaving
the others fixed to their standard values.  The first of these
parameters is related to the amount of kinetic feedback in thick
systems (i.e. in bulges; see point (x) of the list given in
section~\ref{section:morgana}).  As shown in paper I
in an equilibrium condition where the injection of kinetic energy
from SNe is counter-balanced by the dissipation by turbulence,
the velocity
dispersion of cold clouds $\sigma_{\rm cold}$ 
scales with the star-formation timescale $t_\star$ as:

\be \sigma_{\rm cold}=\sigma_0 \left( \frac{t_\star}{1\ {\rm Gyr}}
\right)^{-1/3}\ {\rm km\ s}^{-1}
\label{eq:kinfeed} \ee

\noindent
The normalization parameter $\sigma_0$ depends on many uncertain
details, like the driving scale of turbulence.  In thick systems, due
to the high efficiency of (both thermal and kinetic) energy injection,
$\sigma_0$ is likely to be higher than in thin systems;
this is
observationally demonstrated by Dib, Bell \& Burkert (2006).  In
conjunction with the much lower star-formation time-scale, this can
lead to significant values of $\sigma_{\rm cold}$ in bulges.
The resulting cold wind $\dot{M}_{\rm cw,B}$ is quantified as:

\be
\dot{M}_{\rm cw,B} = M_{\rm c,B} P_{\rm unb}
\frac{v_{\rm unb}}{R_{\rm B}} \label{eq:kinwind_cm}
\ee

\noindent
where $M_{\rm c,B}$ is the bulge cold gas mass, $P_{\rm unb}$ the
probability that a cold cloud with Maxwellian velocity distribution
and rms velocity $\sigma_{\rm cold}$ overtakes the escape velocity of
a bulge with circular velocity $V_{\rm B}$ and $v_{\rm unb}$ is the
average velocity of the unbound clouds.
We will
show in the following that kinetic feedback in bulges plays a very
important role in limiting faint AGNs at high redshift.  We will test
$\sigma_0$ values ranging from 0 to 90 {\kms}.

The second parameter of stellar feedback that is considered here is
$\Sigma_{\rm lim}$, the threshold gas surface density for the switch
to the thick system regime.  This can be explained as follows: as a
result of the strong cooling flows at high redshift, and of the
assumption that the cooled gas settles on a disc, high-redshift discs
may have very high surface densities of cold gas, sufficient to let
them switch to the adiabatic confinement regime, typical of
star-forming thick systems (bulges) and characterized by a higher
velocity dispersion of clouds.  This process is beautifully seen in
some high-redshift starburst galaxies with velocity fields typical of
rotating discs and remarkably high gas velocity dispersions (of order
of $\sim50$ \kms; F\"orster Schreiber et al. 2006).  In these
conditions the transport of angular momentum within the gaseous disc
is more efficient, so that these objects are very likely to eventually
evolve into bulges.  This mechanism is implemented in a very simple
way by stimulating a bar instability (which amounts to moving half of
the disc mass to the bulge) whenever the gas surface density of the
disc overtakes a value $\Sigma_{\rm lim}$; actually, radial flows
typical of bars are observed in some of the F\"orster Schreiber et
al.'s (2006) objects, so this implementation could be realistic.  The
reference value for the $\Sigma_{\rm lim}$ parameter is suggested by
Monaco (2004a; see also paper I) to be 300 \surf; the observations
mentioned above suggest that this may be a conservative choice, but
varying the parameter in the range from 100 to 500 {\surf} does not
lead to very different results.

\subsection{Accretion onto BHs}

This part of the code is described in a simplified way in paper I and
with slight differences in Monaco \& Fontanot (2005).  A seed BH of
$10^3$ {\msun} is assigned to each DM halo, irrespective of its mass
(see, e.g., Volonteri, Haardt \& Madau 2003 for a justification).  Gas
can accrete onto the BH only after having lost nearly all of its
angular momentum $J$; following Granato et al. (2004) we assume that
this low-$J$ gas accumulates in a reservoir of mass $M_{\rm resv}$,
from which it can accrete onto the BH.  The first step in the loss of
angular momentum is connected to the same processes that lead to the
formation of bulges; as a consequence, only the bulge cold phase can
flow in the reservoir.  Further losses of $J$ may be connected to
turbulence, magnetic fields or radiation drag (Umemura 2001); all
these mechanisms are driven by star formation, so the rate of
accumulation of gas into the reservoir, $\dot{M}_{\rm lowJ}$, will be
related to the bulge star-formation rate $\dot{M}_{\rm sf,B}$.  In the
simplest case the two flows will be proportional: $\dot{M}_{\rm lowJ}
= f_{\rm lowJ} \dot{M}_{\rm sf,B}$\footnote{$f_{\rm lowJ}$ was named $k_{\rm
resv}$ in Monaco \& Fontanot 2005.}.  A more general relation between
$\dot{M}_{\rm lowJ}$ and $\dot{M}_{\rm sf,B}$ is obtained assuming a
power-law dependence with exponent $\alpha_{\rm lowJ}$ between the two
quantities:

\be
\dot{M}_{\rm lowJ} = f_{\rm lowJ} \dot{M}_{\rm sf,B} 
\left(\frac{\dot{M}_{\rm sf,B}}{100\ {\rm M}_\odot\, 
{\rm yr}^{-1}}\right)^{\alpha_{\rm lowJ}-1}
\label{eq:lowJ} \ee

\noindent
For $\alpha_{\rm lowJ}=1$ this relation is equivalent to that of
Granato et al. (2004), motivated by the radiation drag mechanism of
Umemura (2001), while for $\alpha_{\rm lowJ}\ne 1$ the $f_{\rm lowJ}$
parameter is scaled to a reference star-formation rate of 100 \msunyr.
For instance, the loss of angular momentum triggered by cloud
encounters will likely have $\alpha_{\rm lowJ}=2$.  A similar approach
is used by Cattaneo et al. (2005)

The gas in the low-$J$ reservoir accretes onto the BH at a rate
determined by the viscosity of the accretion disc; this accretion rate
$\dot{M}_{\rm visc}$ is found by Granato et al. (2004) and in paper I,
but we report it here for completeness:

\be
\dot{M}_{\rm visc}=k_{\rm accr}\frac{\sigma_{\rm B}^3}{G}
\left( \frac{M_{\rm resv}} {M_{\rm BH}} \right)^{3/2}
\left( 1+ \frac{M_{\rm BH}} {M_{\rm resv}} \right)^{1/2}
\label{eq:visc} \ee
 
\noindent
where $\sigma_{\rm B}$ is the 1D velocity dispersion of the bulge and
$k_{\rm accr}$ is theoretically estimated by the authors to be
$\simeq0.001$.  Accretion is limited by the Eddington rate $M_{\rm
BH}/ t_{\rm Ed}$, where $t_{\rm Ed}\simeq 4\times10^7$ yr is the
Eddington-Salpeter timescale.  The resulting system of equations for
the BH mass $M_{\rm BH}$ and reservoir masse $M_{\rm resv}$ is (see also paper I):

\bea
\dot{M}_{\rm BH} &=& \min \left( \dot{M}_{\rm visc}\, , \, \frac{M_{\rm BH}}{t_{\rm Ed}}
\right) \nonumber \\
\dot{M}_{\rm resv} &=& \dot{M}_{\rm lowJ} - \dot{M}_{\rm BH} 
\label{eq:bhaccr} \eea

\subsection{AGN feedback and quasar-triggered winds}
\label{section:qw}

AGN activity releases a huge amount of energy so that, although the
mechanisms for transferring it into the ISM are not very clear, this
energy may easily trigger a massive galactic wind, able to remove all
ISM from the galaxy.  This would mark the end of the star
formation episode that, according to the model described above,
stimulated the accretion onto the BH.  The details of the onset of
such winds are very unclear, so we decide to insert winds in the
model in two ways.

In both cases we use the same criterion for triggering the wind.  This
is motivated in Monaco \& Fontanot (2005) as follows: the UV-X
radiation of the AGN is able to evaporate some 50 \msun\ of cold gas
for each \msun\ of accreted mass.  When this evaporation rate
overtakes the star-formation rate by a factor of order unity 
(0.3 in that paper),
then the
effect of the AGN radiation is sufficient to influence the ISM in a
significant way.  Removing one parameter (the evaporation
efficiency and the triggering parameter are degenerate), the triggering 
condition can be written as:

\bea 
\dot{M}_{\rm BH} > f_{\rm qw} \dot{M}_{\rm sf,B}
\label{eq:trigger1} \eea

\noindent
The quasar-wind parameter $f_{\rm qw}$ determines the critical
accretion rate above which the wind is triggered, and takes values of
order of $10^{-2}-10^{-3}$; 
Monaco \& Fontanot (2005), on the basis of the theoretically
estimated 50 {\msun} of evaporated gas per {\msun} of accreted matter,
used parameter values corresponding to $f_{\rm qw}=0.006$ to obtain
reasonable self-regulated BH masses; we thus use 0.006 as a reasonable
reference value.
Clearly, the criterion of equation~\ref{eq:trigger1} is very similar
to equation~\ref{eq:lowJ} for $\alpha_{\rm lowJ}=1$, which however
refers to the build-up of the reservoir, not to the accretion rate of
the BH.  A modeling of the delay between loss of angular momentum and
accretion onto the BH, dictated by equation~\ref{eq:visc}, is then
necessary to use the criterion of equation~\ref{eq:trigger1}.

A massive removal of cold gas can take place only if the AGN is
powerful enough to perform the work.  Such a self-regulating mechanism
is able by itself to produce a BH -- bulge relation compatible with
the one observed at $z=0$; following again Monaco \& Fontanot (2005)
we put a second condition for the triggering of the wind, requiring
that the mass of cold gas $M_{\rm c,B}$ to be removed from a bulge of
mass $M_{\rm B}$ is not too large:

\be
\frac{M_{\rm c,B}}{M_{\rm B}} <0.21 
\left(\frac{\dot{M}_{\rm BH}}{4\ {\rm M}_\odot\, {\rm yr}^{-1}}\right)^{1.5}
\left(\frac{M_{\rm B}}{10^{11}\ {\rm M}_\odot}\right)^{-1.65}
\label{eq:trigger2} \ee

A third condition for the triggering of a wind is set by requiring the
BH to accrete in a radiatively efficient way; to achieve this we
require that the accretion rate is more than 1 per cent of the
Eddington rate:

\be
\dot{M}_{\rm BH} > 0.01 \frac{M_{\rm BH}}{t_{\rm Ed}}
\label{eq:trigger3} \ee

\noindent
This is motivated by the low radiative efficiency of the flow in case
of low accretion.  These three trigger conditions are those proposed
in the context of the Monaco \& Fontanot (2005) model for the
triggering of the wind, but their validity is wide enough to be used
in a more general context.

The modeling of the wind follows two routes.  As a first option,
``drying winds'' are assumed to remove all the ISM from the bulge
ejecting it to the halo.  This is what happens if the wind is
generated by an injection of kinetic energy coming directly from the
accreting BH.  The ejected gas is assumed to be heated to the
inverse-Compton temperature of the AGN, $T\sim2\times10^7$ K.  Further
accretion is possible from the reservoir, which is not depleted by the
wind.  
This is motivated by the need to have a bright quasar phase after
the wind has removed all the ISM, and may be justified by a bi-polar
outflow that, after piercing the reservoir, generates a blast wave
that becomes symmetric while propagating in the ISM of the inner
galaxy.
As a second option, following the proposal of Monaco \&
Fontanot (2005), ``accreting winds'' are assumed to trigger further
accretion onto the BH.  This is what happens if the wind is generated
throughout the galaxy by SNe\footnote{
In the Monaco \& Fontanot (2005) model the perturbation induced by
runaway radiative heating of cold gas due to the shining of the quasar
is able to trigger a change in the feedback regime that leads to the
creation of a galaxy-wide outflowing cold shell, which is then
effectively pushed out of the galaxy by radiation pressure.
}, then pushed away by radiation pressure of the shining AGN; in this
case a part of the ISM is 
expected
to be compressed to the center, so
that a fraction $f_{\rm centre}$ of the residual cold bulge gas is
given to the reservoir of the BH.  The reference value for this
parameter is set to $0.002$, the case in which $\sim20$ per cent of
the gas is compressed to the centre (as suggested for instance by
Mori, Ferrara \& Madau 2002  in a different context), and 1 per cent
of this gas is able to lose its angular momentum
and flow to the reservoir;
the results are
rather insensitive to the precise value of this parameter.  For
simplicity we neglect any star formation connected to this compressed
gas; this is done to minimize the effect of the highly uncertain
mechanism of quasar winds on the host galaxy.  Clearly, the case
$f_{\rm centre}=0$ corresponds to the drying wind.

Another feedback process that takes place, mostly when BHs are
accreting in a radiatively inefficient regime, is the heating of the
hot halo gas by AGN jets, which can quench the cooling flows in large
DM halos at low redshift.  The importance of this process, which is
complementary to the quasar winds discussed above, is highlighted also
in Fontanot et al. (2006b).  As explained in detail in paper I, we
incorporate a self-consistent implementation of this feedback by
injecting the energy from the accreting BH to the hot halo gas each
time the accretion rate is less than 1 per cent of Eddington; in this
case the radiative efficiency in jets is known to be highest (see,
e.g., Merloni, Heinz \& di Matteo 2003).  For higher 
accretion rates,
we
let only 10 per cent of the emitted energy heat the hot halo gas,
on the ground that $\sim$10 per cent of quasars are radio-loud.  
A remarkable result highlighted in paper I (see its appendix B)
is that the effectiveness of the quenching depends on the mass
resolution of the tree, so that particle masses $\la10^{9}$ {\msun}
are required. However, a proper sampling of the quasar population
requires volumes well in excess of $10^6$ comoving Mpc$^3$; even for
$512^3$ particles this can be achieved only with particle masses
larger than that limit.  We then resort to a simpler recipe for the
quenching of the cooling flow, with a philosophy similar to that of
Bower et al. (2006) and Croton et al. (2006) (see also Cattaneo et
al. 2005 and Kang et al. 2006): we estimate the energy that the BH
accreting at the highest radiatively inefficient rate (i.e. at 0.01
times the Eddington ratio) would give to the hot halo gas,  
then quench the cooling flow each time this energy is higher than that
lost by cooling.  As demonstrated in paper I, this ``forced
quenching'' is able to mimic (with a slight overestimate) the effect
of the self-consistent quenching described above even for poor mass
resolutions.  The mass acquired by the BH in quenching the cooling
flow does not contribute to the AGN LFs, because the accretion is
supposed to be radiatively inefficient by construction, and is not
considered in the BH final mass.  We stress that this ``forced
quenching'' procedure is used in all the models presented below.

\subsection{Runs}
\label{section:runs}

We use the same {\pin} runs introduced in paper I, in particular two
$512^3$ {\pin} realizations of boxes of 200 and 150 comoving Mpc
($h=0.7$), with the cosmological parameters given in the Introduction.
The mass particles are $2.4\times 10^{9}$ and $1.0\times 10^9$ \msun,
so the smallest halo contains 50 particles, for a mass of
$1.2\times10^{11}$ and $5.1\times 10^{10}$ \msun, while the branches
of the DM halo merger trees start at a mass of 10 particles,
corresponding to $2.4\times10^{10}$ and $1.0\times10^{10}$ \msun.  To
sample rare objects like bright quasars a large box is required;
however, as shown in paper I, there is some difference in the results
when passing from the 200 to the 150 Mpc box, most notably in the
amount of star formation at $z>4$ and in the effectiveness of the
self-consistent quenching of the cooling flow (see above,
section~\ref{section:qw}), which is poor for the larger box.  We then
decide to use the large box with the forced quenching procedure
described above, which must be considered as a numerical trick to
mimic the result obtained at higher mass resolution.  As a matter of
fact, the forced quenching procedure is slightly more effective than
the self-consistent one.  In the following we will present only the
results obtained with the 200 Mpc box, but we have 
checked that very similar results (with poorer sampling of the
bright end of the LFs) are obtained with the higher-resolution box and
physical quenching; the details of the quenching procedure have a very
modest influence only on the high-luminosity tail of AGNs at $z<1$.

Considering then the 200 Mpc box, the stellar mass of the typical
galaxy contained in the smallest DM halo at $z=0$ is $\sim 10^{9}$
{\msun}; this is assumed as the completeness limit for the stellar
mass function.  For each run we compute the evolution of (up to) 100
trees (i.e. DM halos at $z=0$) per logarithmic bin of halo mass of
width 0.5 dex.  This implies that while all the most massive halos are
considered, smaller halos are randomly sparse-sampled.  To properly
reconstruct the statistical properties of galaxies we assign to each
tree a weight $w_{\rm tree}$ equal to the inverse of the fraction of
selected DM halos in the mass bin.

The simulated comoving volume of $8\times10^6$ Mpc$^3$ sets an upper
limit $n_{\rm lim}$ to the number density of objects that can be
studied with sufficient statistics.  This limit depends on the
probability of seeing an accretion event (with a given duty cycle) at
a given redshift $z$ in the box, and is computed as:

\be
n_{\rm lim} = \frac{10}{V}\frac{t_{\rm Ed}}{t_{\rm box}(z)}
\label{eq:limit}\ee

\noindent
where the limit refers to 10 objects in the box, $t_{\rm Ed}$ - the
Eddington-Salpeter time - is used as a fiducial duration of an
accretion event, and $t_{\rm box}$ is the cosmological time spanned by
the box at the redshift $z$.  For a box length of 200 Mpc this
function takes values ranging from $10^{-7}$ Mpc$^{-3}$ at $z=0$ to
$2\times10^{-6}$ at $z=5$.  This way we cannot sample the brightest
quasars, characterized by bolometric luminosities well in excess of
$10^{47}$ erg s$^{-1}$.  This allows us to address the bulk of AGN
activity at $0<z<5$, but not to consider the important problem of the
assembly of bright quasars at very high redshift $z\sim6$.  This topic
will be addressed elsewhere, using a larger simulated volume.

\subsection{Computing LFs and X-ray background}

As mentioned in paper I, the information on galaxies is output on a
time grid of 0.1 Gyr.  Due to the short duty cycle of the BH accretion
events, this grid is too coarse to sample properly the AGN activity.
Then, information on BH accretion for all the galaxies is given at
each integration time-step, whenever this accretion is significant; we
use a limit $\dot{M}_{\rm BH}>1.76\times10^{-3}$ \msunyr,
corresponding to bolometric luminosities in excess of $10^{43}$ erg
s$^{-1}$.  With this limit, the detailed information is issued only
for a small fraction of integration time-steps.  In particular, {\gal}
outputs the cosmological time $t$, the integration time-step $\Delta
t$, the BH mass $M_{\rm BH}$, the mass of the reservoir $M_{\rm
resv}$, the bulge total mass $M_{\rm B}$ and cold gas mass $M_{\rm
c,B}$, the accretion rate onto the BH $\dot{M}_{\rm BH}$.  Typically,
each accretion event spans many contiguous timesteps; we treat each
time-step as an independent event with duty cycle $\Delta t$.  The
accretion rate is converted into bolometric luminosity $L_{\rm bolo}$
through the equation $L_{\rm bolo} = \eta \dot{M}_{\rm BH}$, where we
assume that the accreted mass is converted into radiation with an
efficiency of $\eta = 0.1$.  Each event is then counted $w_{\rm tree}$
times to correct for the sampling of the merger trees.

We choose to test our model against observed LFs in the B, soft X-ray
(0.5-2 keV) and hard X-ray (2-10 keV) bands.  The hard X-ray band is
the most useful one to compare with; its main advantage lies in the
low level of extinction suffered by the radiation, especially for
objects at high redshift for which an even harder radiation (in
rest-frame terms) is observed.  In this band we can assume that most
objects are visible, with the only exception of Compton-thick AGNs,
characterized by a hydrogen column density of $N_H>10^{24}$ cm$^{-2}$;
these however may be significant contributors of the X-ray background
(see, e.g., La Franca et al. 2005).  Moreover, as the observed
hardness ratio of a source allows one to estimate $N_H$, it is
possible to estimate the unabsorbed hard-X flux.  We compare our
predictions with analytic fits of the observed LF, corrected for
absorption, proposed by Ueda et al. (2003), Barger et al. (2005) and
La Franca et al. (2005).

Unfortunately, data at such high energies are rather sparse, so it is
useful to compare also with the better sampled soft X-ray band.  The
best available data (Miyaji, Hasinger \& Schmidt 2001; Hasinger,
Miyaji \& Schmidt 2005) are restricted to unabsorbed objects with
$N_H<10^{22}$ cm$^{-2}$, so we need to correct for the fraction of
absorbed objects before comparing model and data.  One possible way
would be to assign an $N_H$ to each AGN event, then selecting only
those that satisfy the selection criterion; this however would
decrease the statistics of the model LF.  We prefer then to compute
the fraction of unabsorbed objects in luminosity bins, given the $N_H$
distribution, and correct our LF by that fraction.  For the $N_H$
distribution we use the luminosity- and redshift-dependent one
proposed by La Franca et al. (2005).

We also compare our model to the B-band LFs (Kennefick, Djorgovski \&
Meylan 1996; Fan et al. 2003; Croom et al., 2004; Wolf et al. 2003;
Cristiani et al. 2004; Fontanot et al. 2006a), that are measured with
the best statistics and in the widest redshift range.  As a matter of
fact, the most stringent constraint comes from the high-redshift
($3.5<z<5.2$), low-luminosity ($M_{\rm B}\sim -22$ to $-24$) AGNs observed
by GOODS (Cristiani et al. 2004; Fontanot et al. 2006a).  
To compute our predicted B-band LF of type-I objects we assume
that the type I fraction depends on luminosity according to the
correlation found by Simpson (2005).

To transform from bolometric to band luminosities we use the Elvis et
al. (1994) bolometric correction for the B-band, assuming a value of
$10.4 \pm 2.0$ for the ratio $\nu_{4400} f_{\nu,4400}$ and the
bolometric luminosity $L_{\rm bolo}$.  In the X-ray bands we adopt the
bolometric corrections proposed by Marconi et al. (2004):

\bea
Log(\frac{L_{\rm bolo}}{L_{\rm soft}}) = 1.64 + 0.22 L_{12} + 0.012 L_{12}^2 -
0.0015 L_{12}^3 \nonumber\\
Log(\frac{L_{\rm bolo}}{L_{\rm hard}}) = 1.53 + 0.24 L_{12} + 0.012 L_{12}^2 -
0.0015 L_{12}^3 \label{eq:marconi}
\eea

\noindent
where $L_{12} = Log(L_{\rm bolo})-12$.  Marconi et al. (2004) also propose
a luminosity-dependent bolometric correction for the B-band which is
in agreement with the Elvis et al. correction in the $L_{\rm bolo}$ range
of our interest.

As a consistency check, we compare our models also to number counts
and the X-ray background.  To compute our predictions for these
quantities we build a library of template spectral energy
distributions (SEDs) as follows (see also Monaco \& Fontanot 2005;
Ballo et al. 2006).  In the optical we consider the quasar template
spectrum of Cristiani \& Vio (1990) down to 538 \AA, extrapolated to
300 \AA\ using $f_\nu \propto \nu^{-1.76}$ (following Risaliti \&
Elvis 2005).  At shorter wavelengths (between 0.01 and 30 \AA) we use
a power-law SED with a photon index of $\Gamma=-1.8$ and an
exponential cutoff $\exp(-E/200 {\rm keV})$ (see i.e. La Franca et
al., 2005).  The relative normalization between the optical-UV and the
X-ray branches of the spectrum is constrained through the quantity
$\alpha_{ox}$ (Zamorani et al., 1981):

\be
\alpha_{ox} = \frac{log(L_{\nu_{2500}} / L_{\nu_{1 {\rm keV}}})} 
{log(\nu_{2500} / \nu_{1 {\rm keV}})}
\ee

\noindent
We use for $\alpha_{ox}$ the bolometric luminosity-dependent of
$\alpha_{ox}$ value proposed by Vignali, Brandt \& Schneider (2003):

\be
\alpha_{ox} = -0.11 \, log(L_{\nu 2500}) +1.85
\ee

\noindent
The interpolation between 30 and 300 \AA\ follows Kriss et al. (1999).
We then produce a library of template spectra in the range ${\rm Log}(
L_{\rm bolo} )= [42.0,47.5]$ (divided into bins of 0.1 dex in
luminosity).  For each accretion event in our {\gal} output we
associate a template spectrum and an $N_H$ value, extracted from the
La Franca et al. (2005) distribution, which includes also
Compton-thick objects (for which the X-ray flux is set to zero).  The
$N_H$ absorption is computed using the Morrison \& McCammon (1983)
cross section.  We also compute absorption by the ISM following Madau,
Haardt \& Rees (1999); this is important only at the lowest energies.
The integration in redshift is easily performed with the {\gal}
output, as this spans the whole range of cosmological times from
recombination to the present.  This is at variance with Fontanot et
al. (2006b), where the computation of the galaxy SEDs is much more
demanding in terms of computer time, so that a very careful sampling
of model galaxies is needed.

\subsection{Parameter space and models}
\label{section:parspace}

Starting from the standard set of parameters defined in paper I, we
have investigated a limited subset of parameters that influence
significantly the AGN population but are varied in a limited range so
as to give very similar results in terms of galaxies.  These are the
galaxy feedback parameters $\sigma_0$ (which sets the level of kinetic
feedback) and $\Sigma_{\rm lim}$ (which sets the feedback-induced bar
instability), and the quasar wind parameters $f_{\rm qw}$ (which sets
the trigger for quasar winds), $f_{\rm lowJ}$ (which regulates the
quantity of gas flowing into the reservoir), $\alpha_{\rm lowJ}$
(which sets the scaling of the angular momentum loss with the bulge
star-formation rate) and $f_{\rm centre}$ (which sets the fraction of
bulge gas that is available for accretion after the wind is
triggered).

\begin{table*}
\begin{center}
\begin{tabular}{l|llllll}
Model & $\sigma_0$ & $\Sigma_{\rm lim}$ & $f_{\rm qw}$ & $f_{\rm lowJ}$ & 
$\alpha_{\rm lowJ}$ & $f_{\rm centre}$ \\ \hline
STD   & 60 \kms & $\infty$ \surf & 0     & 0.003 & 1 & 0 \\
DW    & 60 \kms & $\infty$ \surf & 0.006 & 0.03  & 2 & 0 \\
AW    & 60 \kms & 300 \surf      & 0.006 & 0.01  & 1 & 0.002 \\
\end{tabular}
\end{center}
\caption{Parameter values for the models used in this paper.}
\label{table:mod}
\end{table*}

We have investigated this parameter space, finding more that one
acceptable solution.  Here we present the results obtained with three
models, namely the standard model presented in paper I (called STD), a
model with drying winds and $\alpha_{\rm lowJ}=2$
(DW) and a model
with accreting winds $\Sigma_{\rm lim}=300$ {\surf} (AW).  The
relative parameters are reported in table~\ref{table:mod}.  It is
worth noticing that the two models with quasar winds have higher
$f_{\rm lowJ}$ values:
in the STD model $f_{\rm lowJ}$ sets the the level of the BH --
bulge relation, while in the wind models DW and AW the final BH mass
is not set by $f_{\rm lowJ}$ but by the self-regulating action of the
wind; in this case much higher $f_{\rm lowJ}$ values can be allowed,
and this parameter regulates the early bulding of the massive BHs and
then the bright end of the AGN LFs.


\section{Results}
\label{section:results}

\begin{figure*}
\centerline{
\includegraphics[width=14cm]{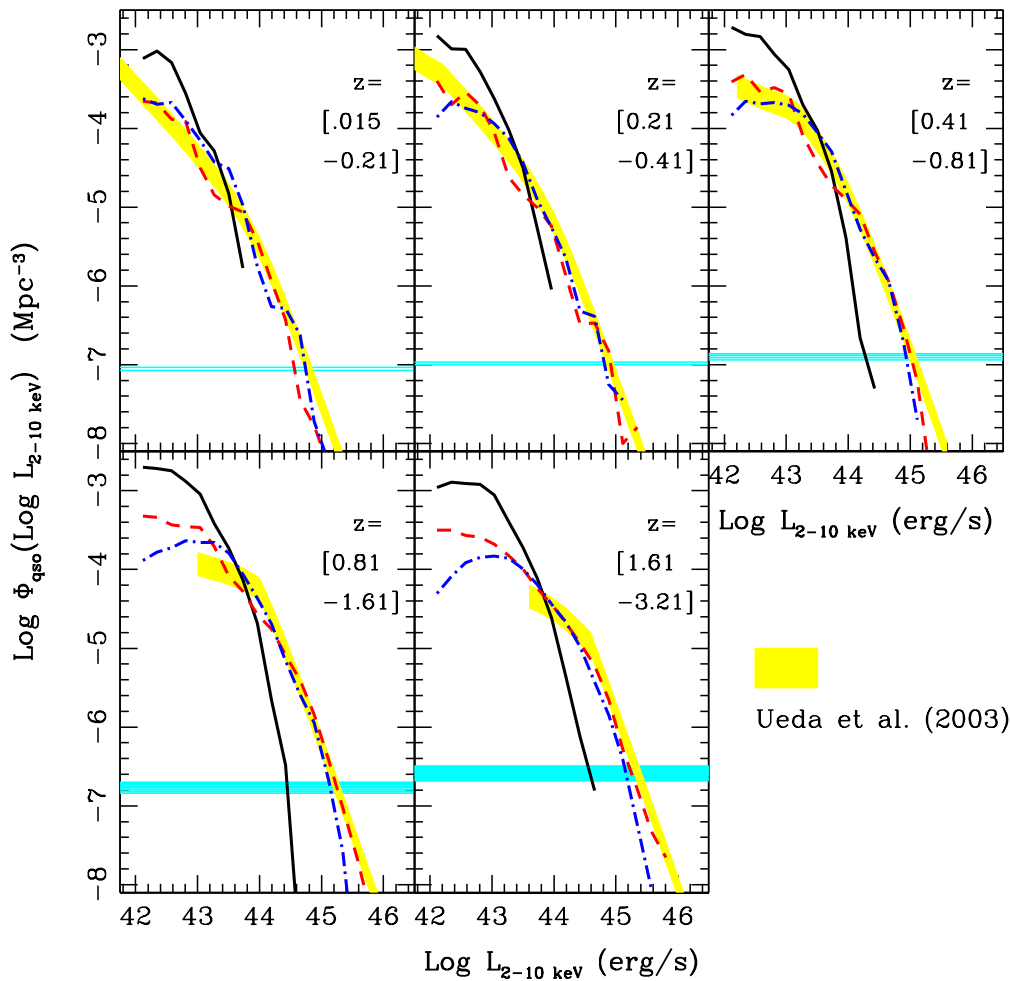}
}
\caption{2-10 keV X-ray LF. The lines are the predictions of the STD
model (black solid line), DW model (red dashed line), AW model (blue
dot-dashed line).  The shaded region represents the Ueda et al (2003)
estimate.  The cyan horizontal bands give the limit below which the LF
is affected by poor statistics in our run (equation~\ref{eq:limit})}
\label{fig:hx_lf}
\end{figure*}

\begin{figure*}
\centerline{
\includegraphics[width=14cm]{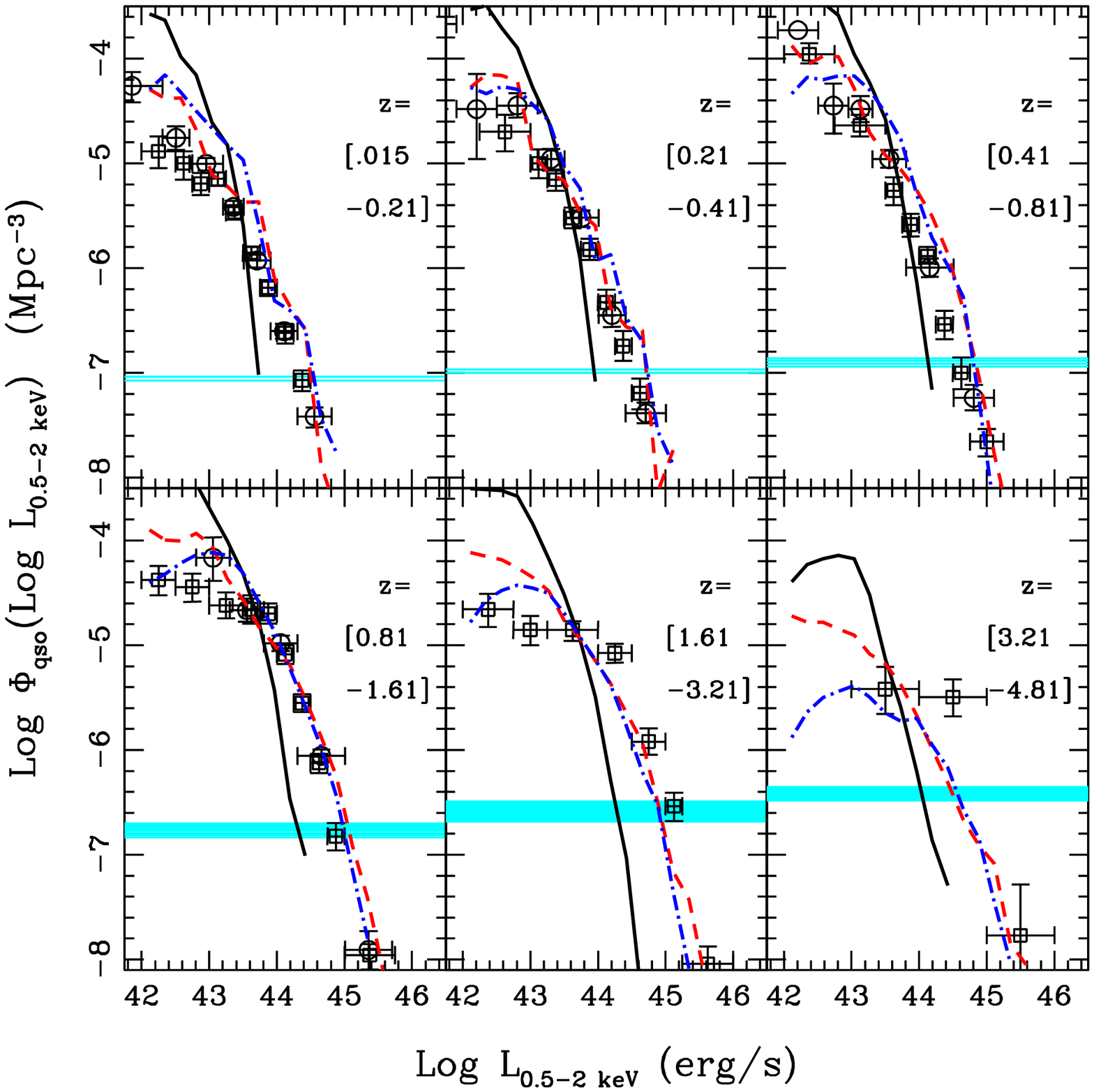}
}
\caption{0.5-2 keV X-ray LF.  Lines refer to models as in
figure~\ref{fig:hx_lf}. In all panels the empty circles and squares
refer respectively to the observations of Miyaji et al. (2001) and
Hasinger et al. (2005).}
\label{fig:sx_lf}
\end{figure*}

\begin{figure*}
\centerline{
\includegraphics[width=14cm]{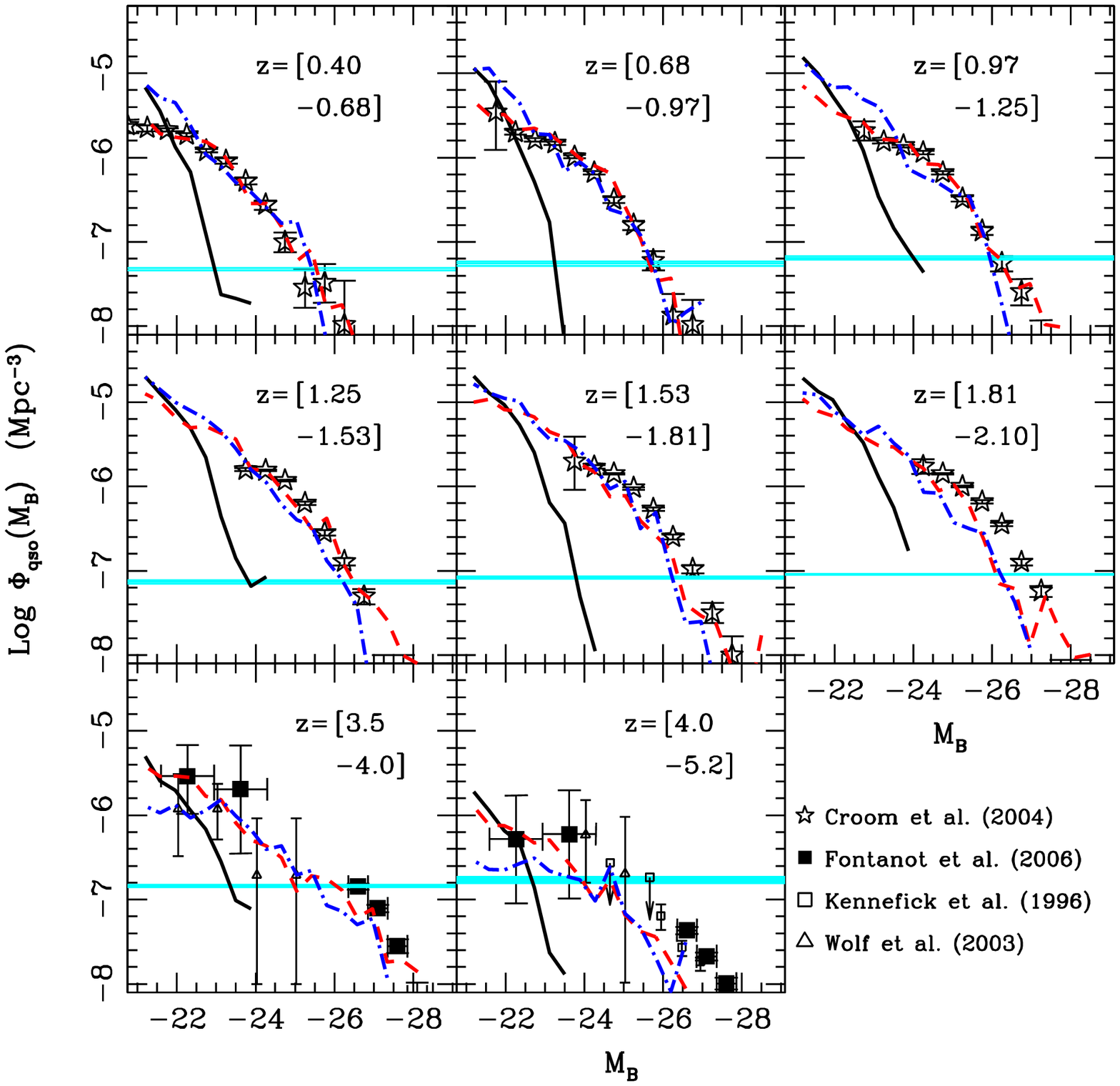}
}
\caption{B-band LF.  Lines refer to models as in
figure~\ref{fig:hx_lf}.  Data are taken from Kennefick et al. (1996),
Wolf et al. (2003), Croom et al. (2004), Fontanot et al. (2006a), as
indicated in the figure.}
\label{fig:opt_lf}
\end{figure*}

\begin{figure*}
\centerline{
\includegraphics[width=14cm]{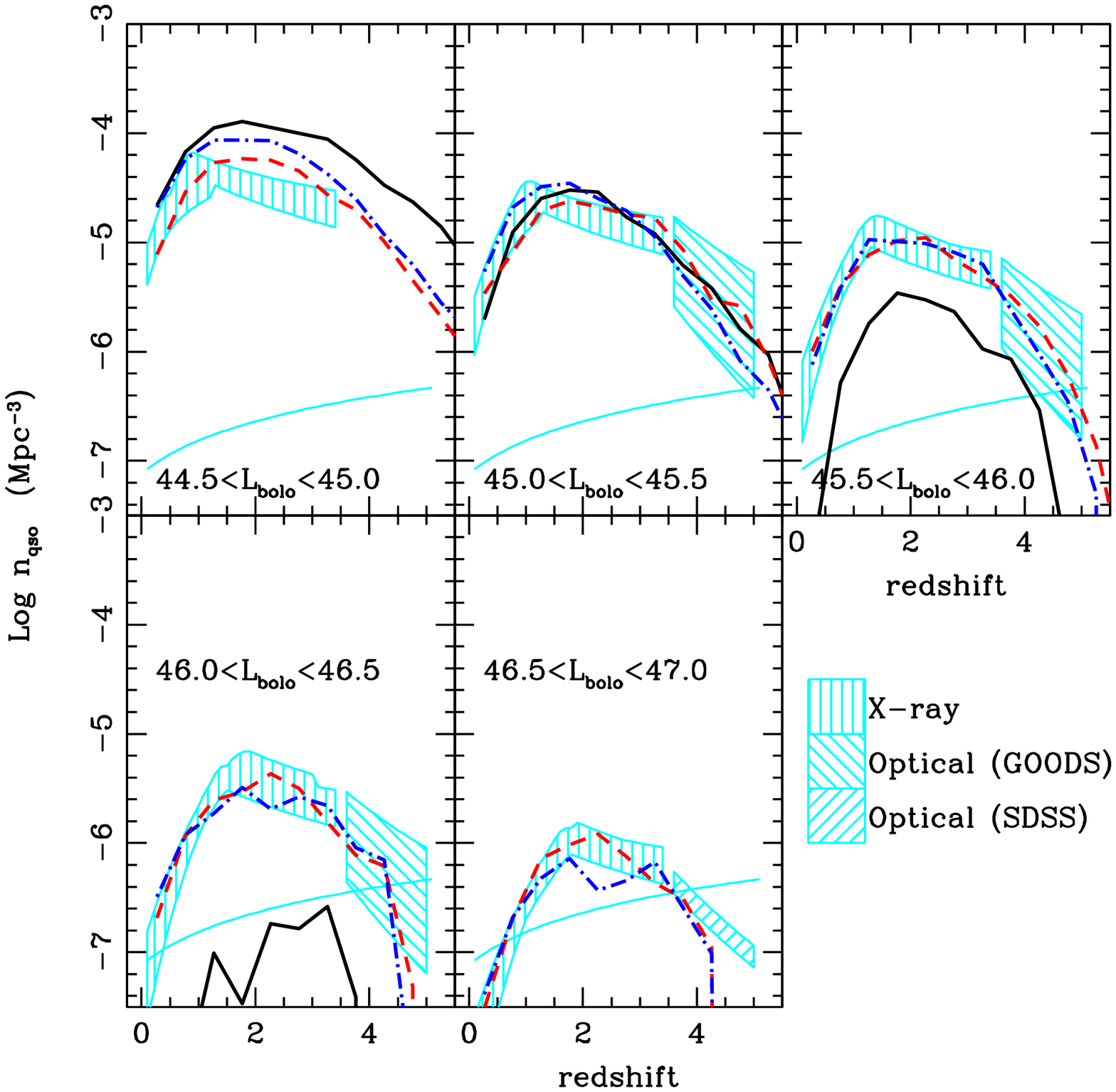}
}
\caption{Evolution of the QSO number density with redshift in
different bins of bolometric luminosity, estimated from X-ray (Ueda et
al. 2003; Barger et al. 2003; La Franca et al. 2005) and optical
(SDSS, Fan et al. 2003; GOODS, Fontanot et al. 2006a) surveys. Lines
refer to models as in figure~\ref{fig:hx_lf}.}
\label{fig:numdens}
\end{figure*}

\begin{figure}
\centerline{
\includegraphics[width=9cm]{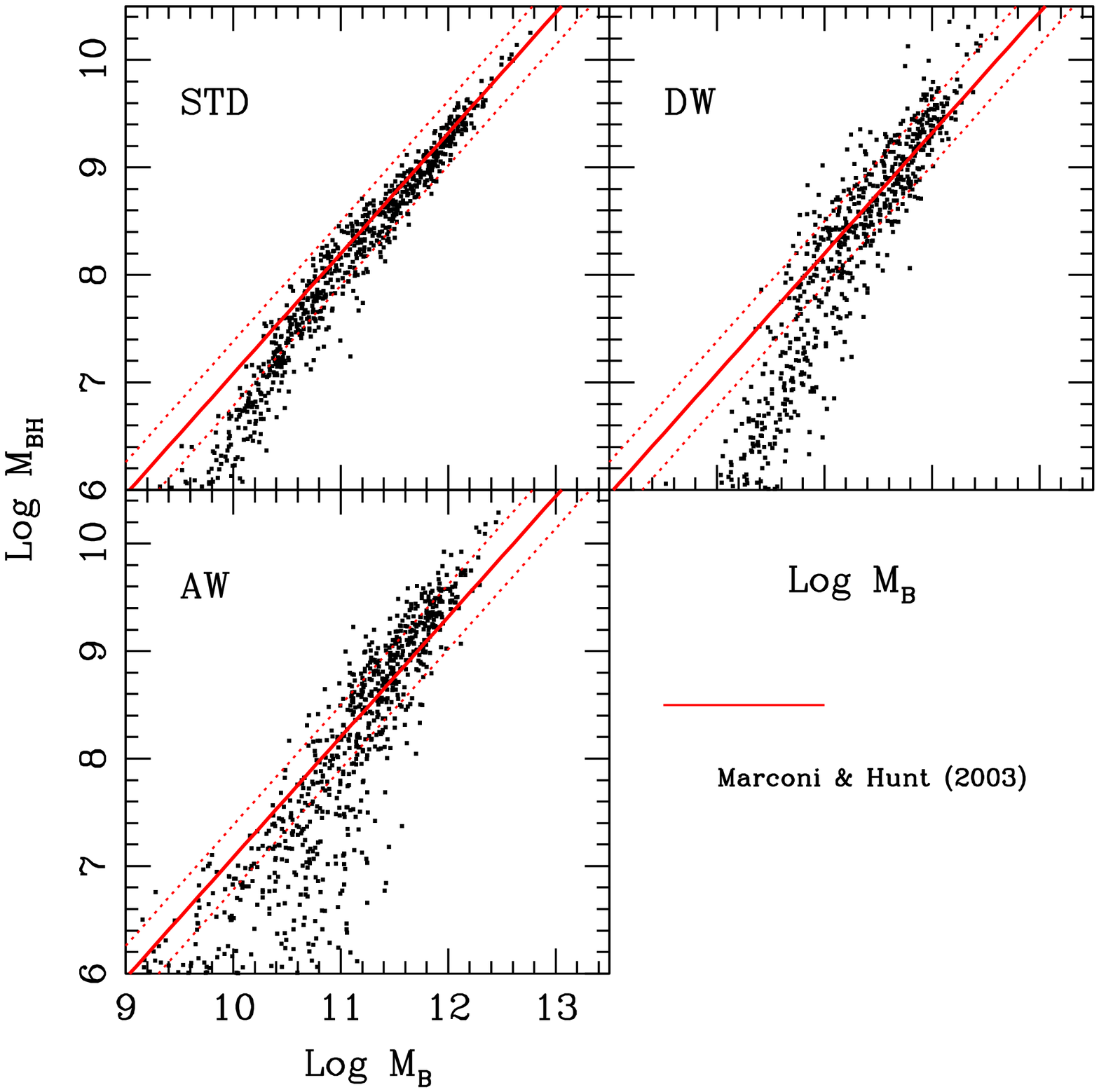}
}
\caption{Predicted $M_{\rm B}-M_{\rm BH}$ relation at $z=0$ for STD, DW and
AW models.  The thick diagonal line gives the observed mean relation
according to Marconi \& Hunt (2003), the dotted lines highlight the
observed scatter of 0.3 dex.}
\label{fig:bhbulz0}
\end{figure}

In this section we compare the results of our model to observations of
LFs and number counts in the hard-X, soft-X and optical bands, and to
the statistics of remnant BHs at $z=0$.  To best illustrate the
constraints that can be obtained on the physical processes described
above we show results for the three combination of parameters given in
table~\ref{table:mod} and introduced in
section~\ref{section:parspace}.  The STD model refers to the standard
choice of parameters presented in paper I, with the addition of the
forced quenching procedure described in section~\ref{section:qw}.  In
this case quasar winds are not active ($f_{\rm qw}=0$).  The resulting
LFs for this model are always too steep
(figures~\ref{fig:hx_lf}--\ref{fig:numdens}) though the $z=0$ BH --
bulge relation is roughly reproduced (figure~\ref{fig:bhbulz0}) for
$M_{\rm B}\ga10^{11}$ \msun.  Within the range of parameters allowed
by the constraints of galactic observables (see paper I and Fontanot
et al. 2006b) we have found no way to have shallower LFs.  On the
other hand, the introduction of quasar winds allows us to improve the
agreement with AGN data without influencing much the results on
galaxies.  The DW model includes drying winds as follows.  In order to
flatten the AGN LF the $\alpha_{\rm lowJ}$ exponent is set to 2, so
that accretion is increased in the strong starbursts that give rise to
big spheroids, and depressed in small bulges; in order to avoid a
dramatic steepening of the BH -- bulge relation at $z=0$, drying winds
are introduced to limit the mass of the most massive BHs.  In this
case we use $f_{\rm lowJ}=0.03$ in place of the 0.003 value of STD.
Another good combination of parameters (model AW) is found by allowing
accreting winds (with $f_{\rm qw}=0.006$, $f_{\rm lowJ}=0.01$ and
$f_{\rm centre}=0.002$) and setting $\Sigma_{\rm lim}=300$ \surf\
(model AW); in this case both the starbursts induced in discs with a
high gas surface density and the secondary accreting episodes
stimulated by the quasar wind increase the activity of bright quasars
even with $\alpha_{\rm lowJ}=1$,

Figures~\ref{fig:hx_lf}, \ref{fig:sx_lf} and \ref{fig:opt_lf} show
that the DW and AW models fit nicely (though maybe not in detail) the
observed LFs of quasars in all the bands considered.

Figure~\ref{fig:numdens} shows the same results in a different way.
The predicted number density of AGNs in bins of bolometric luminosity
is compared to the range of values inferred from observations (using
mainly the analytic fit of the hard-X LFs of Ueda et al. 2003, Barger
et al. 2003 and La Franca et al. 2005 for $z<3.5$, and the results of
Fontanot et al. 2006a at $z>3.5$).  The three models are all shown.
It is clear that the DW and AW models reproduce nicely the evolution
of the AGN population, with the number distributions peaking at
lower redshift for lower luminosity.  In other words, the downsizing
or anti-hierarchical behavior of AGNs can be recovered in the context
of the hierarchical $\Lambda$CDM cosmogony once feedback is properly
modeled.  

Similar results are obtained with the 150 Mpc box, though the
statistical limit for the number density of objects
(equation~\ref{eq:limit}) is higher by a factor of 2.4, the decrease
of the AGN activity at low redshift is marginally slower and the
slight excess of faint AGNs at $z\sim 2$ is a bit larger.  Clearly,
some of these discrepancies can be absorbed by a further tuning of the
parameters, which we deem worthless in this analysis.

\begin{figure*}
\centerline{
\includegraphics[width=14cm]{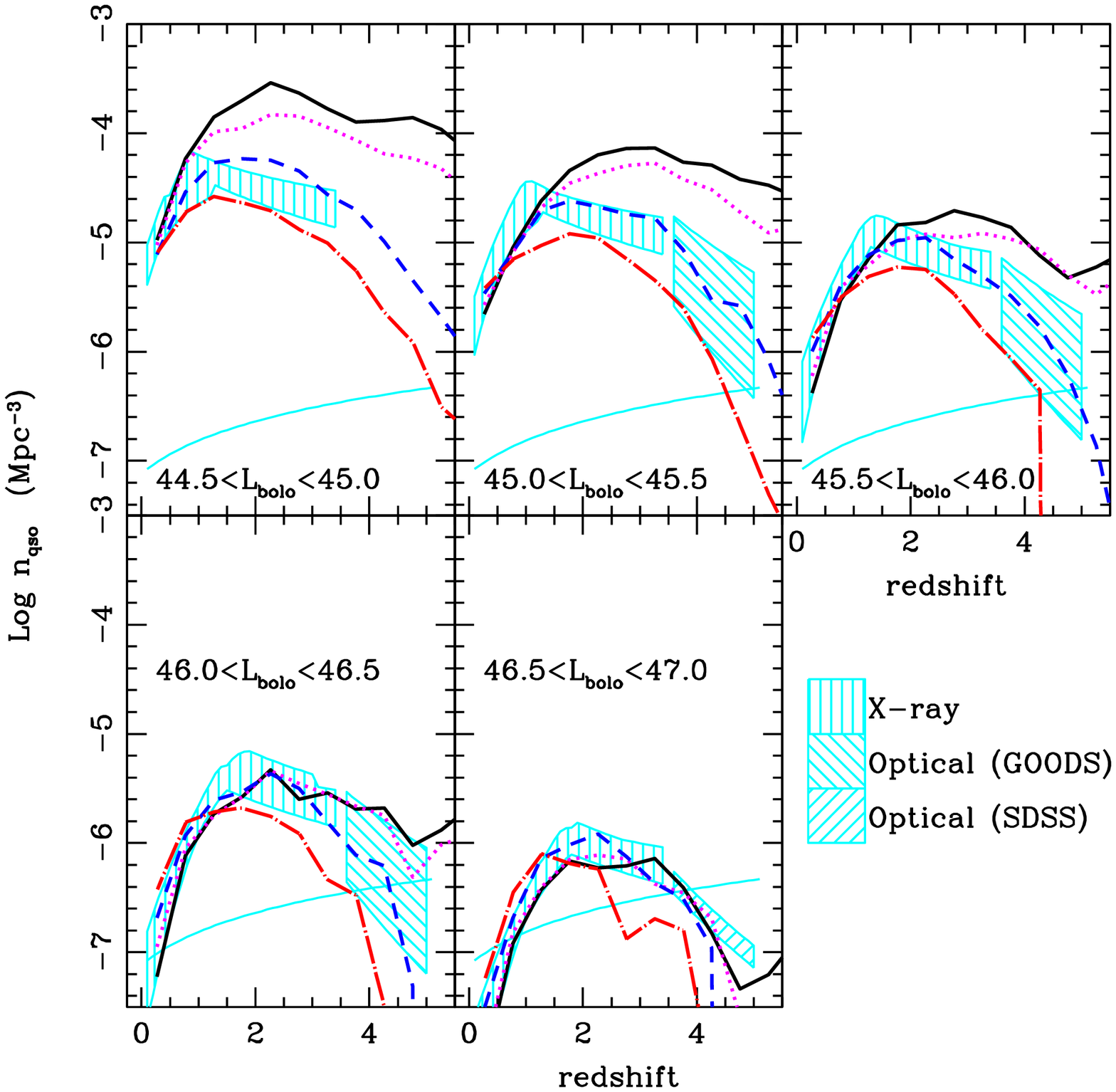}
}
\caption{Evolution of the QSO number density for the DW model as a
function of $\sigma_0$.  The black solid line refers to $\sigma_0=0$
\kms, the magenta dotted line to $\sigma_0=30$ \kms, the blue dashed
line to $\sigma_0=60$ {\kms} (the best-fit value) , the red dot-dashed
lines to $\sigma_0=90$ \kms.}
\label{fig:v_turb}
\end{figure*}

From a much more detailed analysis of the parameter space we infer
that the physical process at the origin of the downsizing of AGNs is
the kinetic feedback active in star-forming bulges.  This is shown in
figure~\ref{fig:v_turb}, where the DW model is shown with $\sigma_0$
values ranging from 0 to 90 {\kms} (very similar results are obtained
with the other models). Kinetic feedback causes a strong ejection of
cold gas in small bulges, thus decreasing the number of faint AGNs
without changing much the number of bright quasars.  From this
comparison we obtain a best-fit value for $\sigma_0$ of 60 \kms; this
is the most effective way to constrain this parameter.  It is
also worth noticing in figure~\ref{fig:v_turb} the fundamental
importance of the constraint coming from the abundance of high-$z$
AGNs in the GOODS fields (Cristiani et al. 2004; Fontanot et
al. 2006a).

\begin{figure}
\centerline{
\includegraphics[width=9cm]{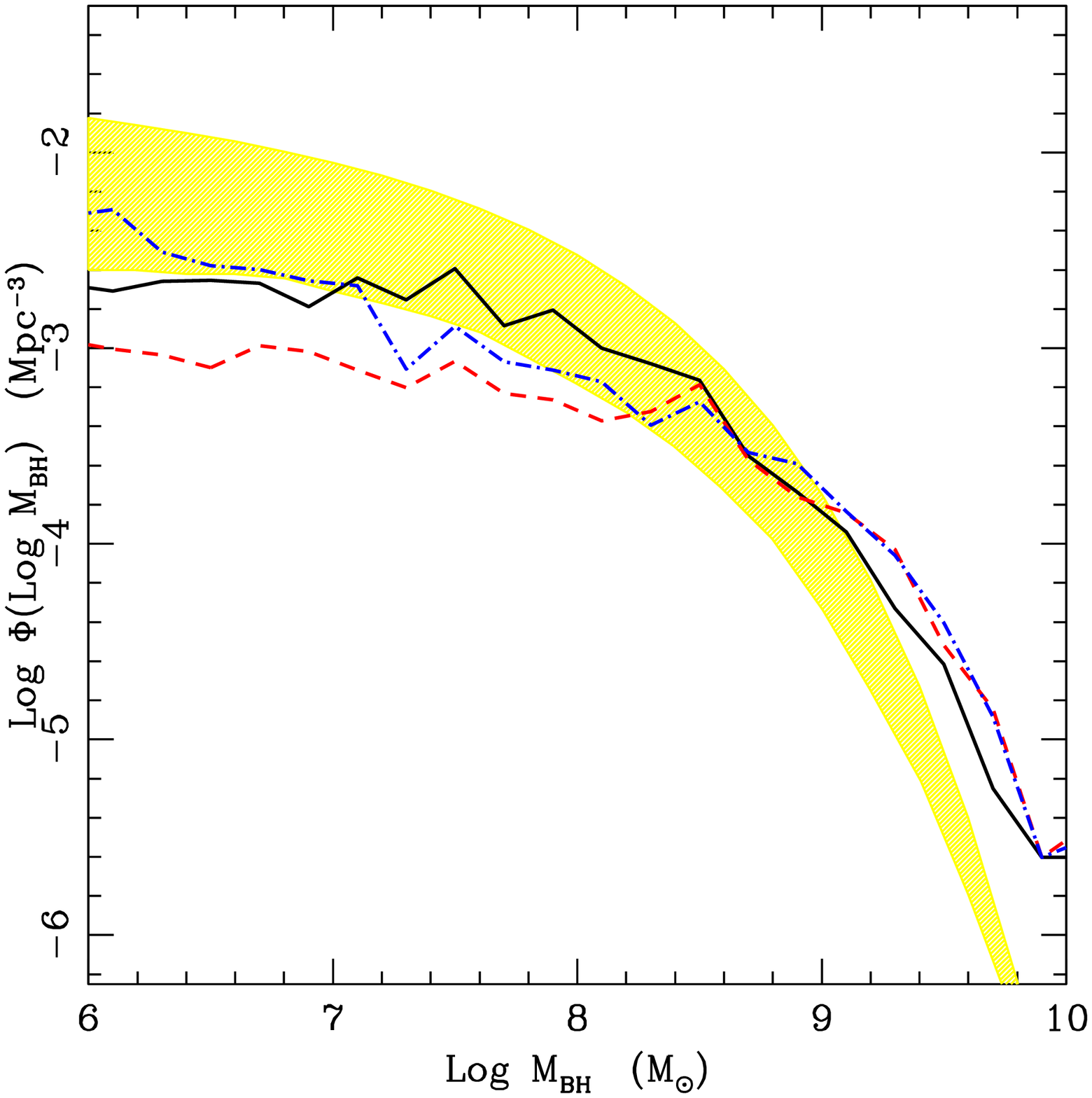}
}
\caption{BH mass function at $z=0$ for the models.  (lines refer to
 models as in figure~\ref{fig:hx_lf}), compared to the region allowed
 by the Shankar et al. (2004) and Marconi et al. (2004) estimates
 (shaded area).  }
\label{fig:mf_bh}
\end{figure}

Figure~\ref{fig:bhbulz0} shows the BH -- bulge relation at $z=0$
predicted by the models, while figure~\ref{fig:mf_bh} shows the
resulting mass function of BHs.  All models, included STD, roughly
reproduce the observed range of values for massive bulges ($M_{\rm
B}\ga10^{11}$ \msun), with some modest overestimate by the wind models
(especially AW), while a steepening is predicted at smaller
masses.  This figure allows to make some important points. First,
despite its fundamental importance in demonstrating the connection
between AGNs and their host bulges, this relation is not after all a
very strong constraint, as the three models give similar fits (for
$M_{\rm B}>10^{11}$ \msun) despite their remarkably different
performance in terms of LFs and number densities of objects.  Second,
the scatter of this relation is remarkably low in the STD model, and
this suggests that the recipes given in section~\ref{section:qw} are
too simple and some source of scatter is needed.  Intriguingly, quasar
winds give roughly the correct amount of scatter, though they tend to
give too many massive BHs at $z=0$.  Third, BHs in small bulges are too
small compared to observations.  This very important point will be
deepened below.

\subsection{X-ray number counts and background}
\label{section:xrb}

\begin{figure}
\centerline{ 
\includegraphics[width=9cm]{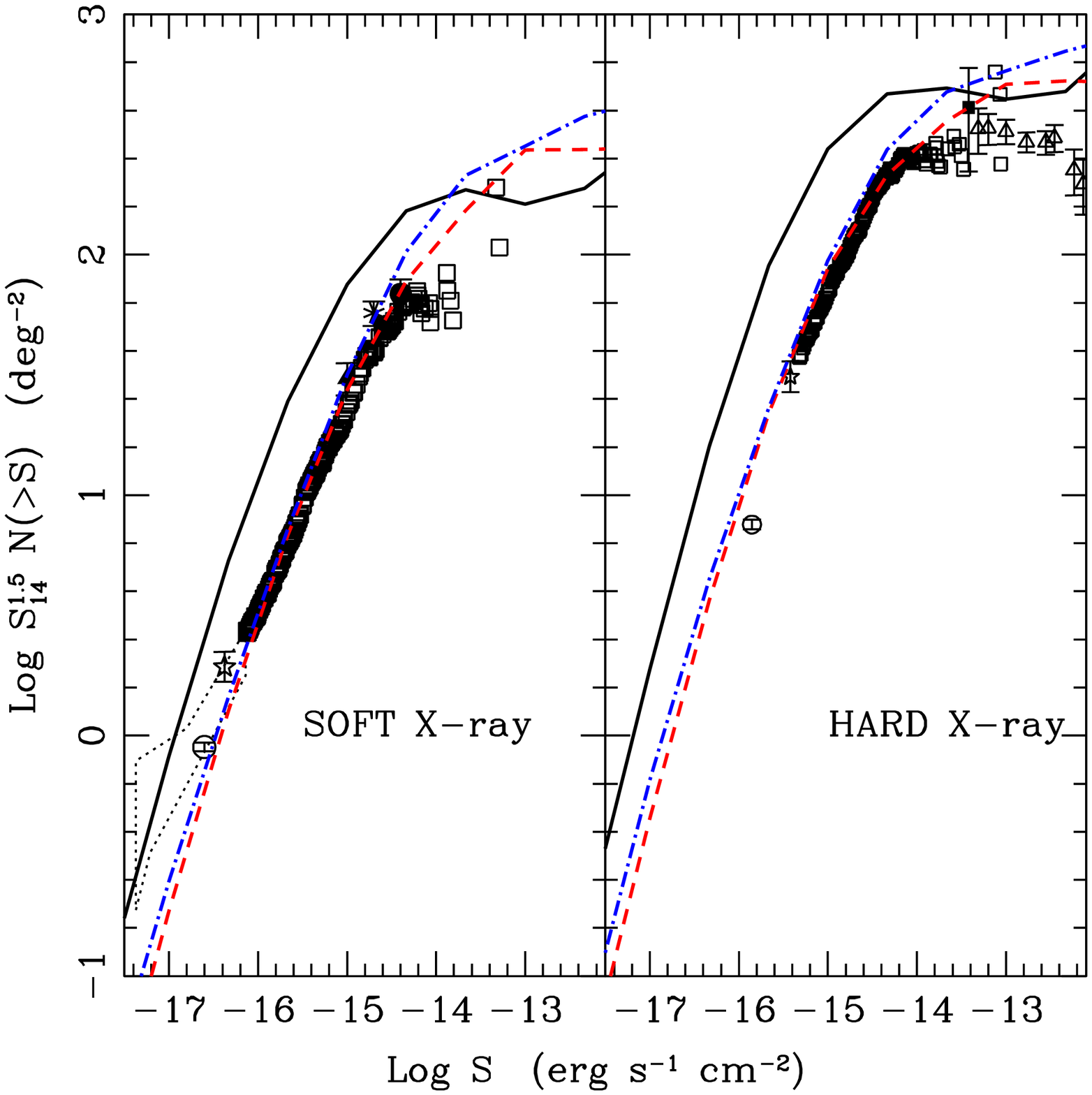} 
}
\caption{Cumulative Source Number Counts in X-ray bands. Left Panel:
soft (0.5-2 keV).  Estimates from Hasinger et al., (1993) ({\it dotted
region}); Brandt et al., (2001) ({\it empty star}); Bauer et al.,
(2004) ({\it empty circle}); Hasinger (1998) ({\it filled triangle});
Zamorani, Mignoli \& Hasinger (1999) ({\it filled circle}); Rosati et
al. (2002) ({\it empty squares}). Right Panel: hard (2-10
keV). Estimates from Giommi et al., (2001) ({\it empty triangles});
Ogasaka et al. (1998) ({\it filled square}); Brandt et al., (2001)
({\it empty star}); Bauer et al., (2004) ({\it empty circle}); Rosati
et al. (2002) ({\it green empty squares}). Lines refer to models as in
figure~\ref{fig:hx_lf}}
\label{fig:xnc}
\end{figure}

In order to strengthen the constraints on the AGN population we
compare our model predictions to X-ray number counts
(figure~\ref{fig:xnc}) in the soft and hard X-ray bands.  Apart from
some possible overestimate at the brightest fluxes, the fit to the
data of the two wind models (DW and AW) is very good, while the STD
model overestimates the counts by roughly a factor of three, and this
is due to the overestimate of the faint end of the LFs.

\begin{figure}
\centerline{
\includegraphics[width=9cm]{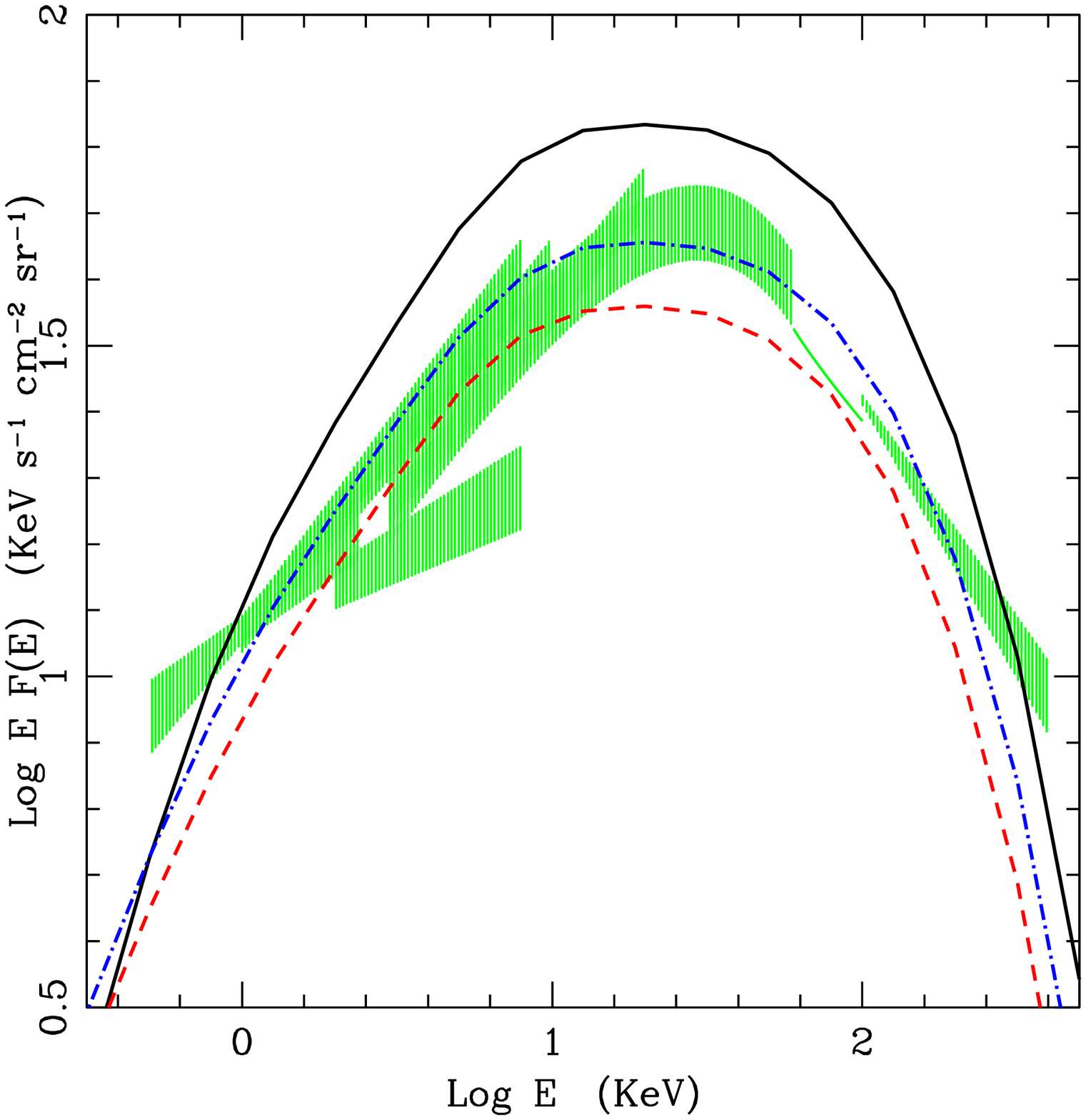}
}
\caption{Reproduction of the X-ray background. Observations are
highlighted by the shaded areas and are taken from Revnivtsev et
al. (2003), Worsley et al. (2004), Georgantopoulos et al (1996), Vecchi
et al. (1999), Lumb et al., (2002), De Luca \& Molendi (2004), Kinzer et
al., (1997), Gruber et al. (1992). Lines refer to models as in
figure~\ref{fig:hx_lf}}
\label{fig:xrb}
\end{figure}

Figure~\ref{fig:xrb} shows the prediction for the X-ray background
from 0.5 to 300 keV.  The background predicted by the two wind models
follows nicely the observed one, though the peak at $\sim30$ keV is
underpredicted by the DW model.  This is in line with the results of
La Franca et al. (2005), who claim a missing (though not dominant)
population of Compton-thick AGNs, which is not included in this
background synthesis 
that uses similar ingredients as their paper.
This agreement shows that also the population of faint sources is
roughly reproduced by the wind models.

\subsection{More predictions}
\label{section:predictions}

\begin{figure}
\centerline{
\includegraphics[width=9cm]{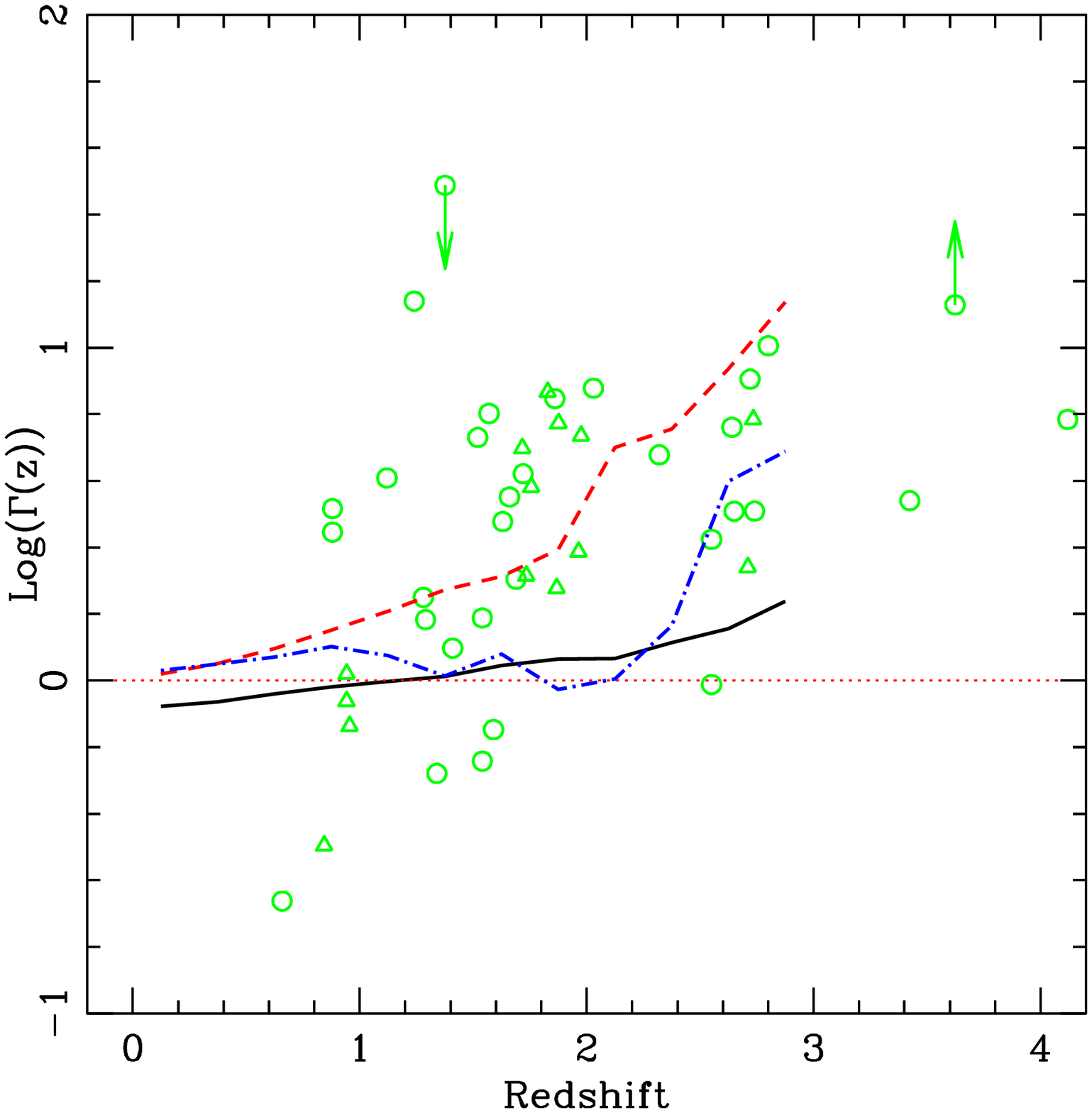}
}
\caption{Evolution in redshift of the BH--bulge relation through the
$\Gamma$ parameter (equation~\ref{eq:peng}) for the three models.
Lines refer to models as in figure~\ref{fig:hx_lf}.  Data points are
taken from Peng et al. (2006).}
\label{fig:peng}
\end{figure}

In order to quantify the amount of evolution in the BH-bulge
relation we compare our model predictions with the results of Peng et
al. (2006), who measured the evolution of the BH -- bulge relation by
defining the quantity $\Gamma(z)$:

\be
\Gamma(z) = \frac{M_{\rm B}}{M_{\rm BH}}(z)/ \frac{M_{\rm B}}{M_{\rm BH}}(z=0)
\label{eq:peng}
\ee

In figure~\ref{fig:peng} we present the average evolution of $\Gamma$
predicted by the three models, but only for the galaxies with $M_{\rm
B} > 2 \times10^{11}$, which safely lie on the relation.  The model
STD show some degree of evolution, which is due to the fact that,
consistently with Croton (2006), the fraction of stars formed in
bulges is higher at higher redshift, so that a higher fraction of gas
can accrete on the BH.  On the other hand, the wind models show a more
pronounced increase of $\Gamma$ with redshift, in better agreement
with Peng et al. (2006).  This was anticipated in Monaco \& Fontanot
(2005), who showed that if more mass is allowed to flow onto the BH
until the winds limits further accretion, then the mass assembly of
massive BHs is anticipated.  This can help in reconciling the presence
of bright quasars at high redshift, when very massive galaxies were
not yet assembled.

\begin{figure}
\centerline{
\includegraphics[width=9cm]{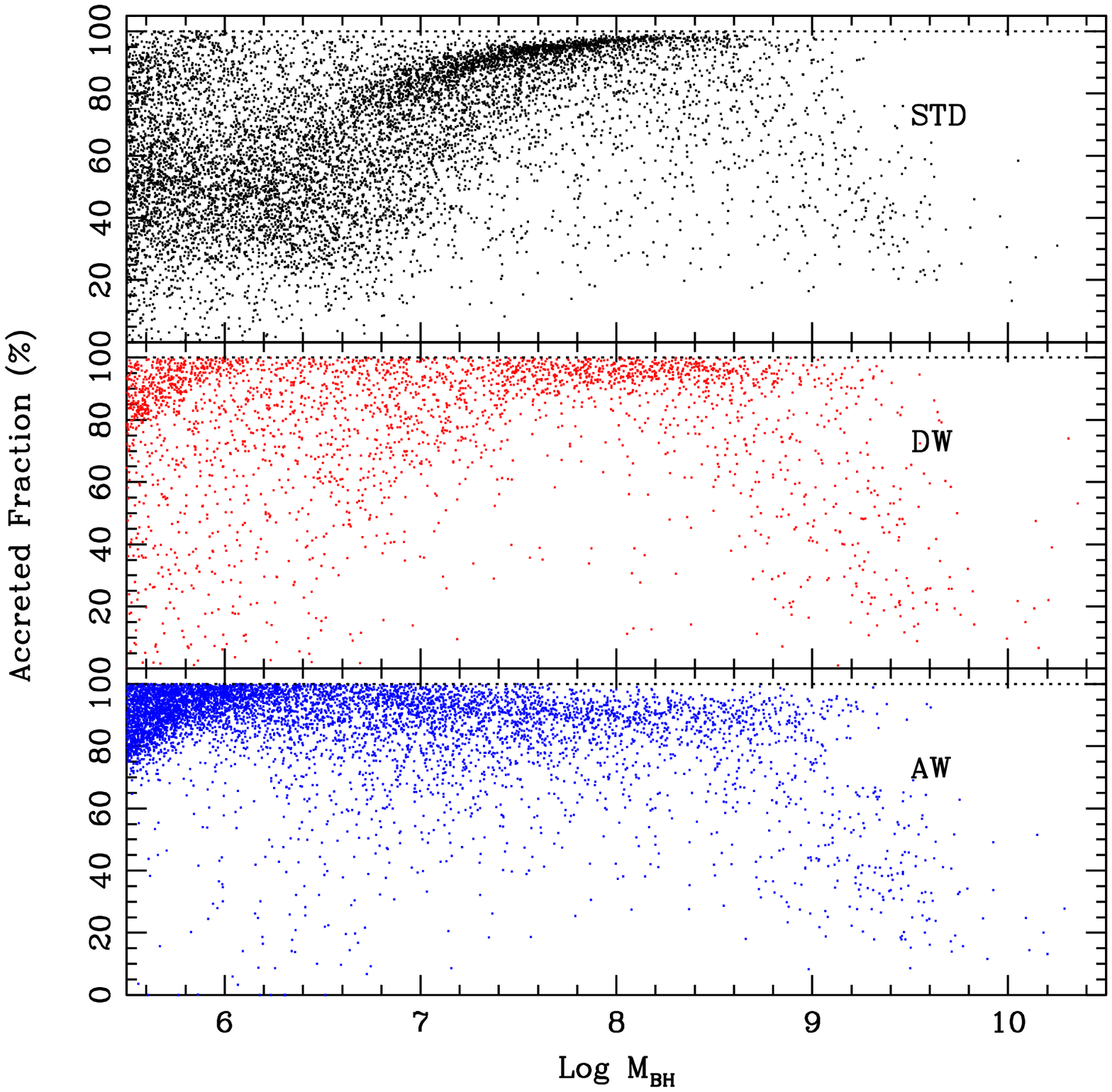}
}
\caption{Scatter-plot of the predicted fraction of BH mass acquired by
accretion as a function of BH mass, for BHs in $z=0$ galaxies.  The
three panels give the results of the STD (upper), DW (mid) and AW
(lower) model.}
\label{fig:acc_frac}
\end{figure}

Regarding the mechanisms responsible for the assembly of BHs, we
find that the bulk of BH mass is gained from accretion.
Figure~\ref{fig:acc_frac} shows the fraction of mass acquired by
accretion (the rest is acquired by mergers) as a function of BH mass
for the BHs found in $z=0$ galaxies, according to the three models.
We notice that merging is significant for the most massive BHs, which
reside in the most massive galaxies that experience complex merger
histories (see, e.g., De Lucia et al. 2006; De Lucia \& Blaizot
2006). We also notice that quasar winds, especially the accreting
winds, increase the amount of mass acquired by accretion.

Another important prediction is the average accretion rate of BHs in
units of the Eddington rate.  This quantity is needed to relate the
accretion history of AGNs, estimated from the LFs, to the local BH
mass function.  As each accretion event is subdivided by our model
into many sub-events, one per integration interval, we show this
quantity averaged over bins of bolometric luminosity and redshift and
weighted by the width of the time bin.  Figure~\ref{fig:ed_lum} shows
the results for the STD, DW and AW models.  While the STD rates are
very low, especially for the highest accretion rates, the Eddington
ratios of the other models are rather high at high redshift and
decrease at low redshift, especially for low AGN luminosities.
This prediction is compared with the observational results of
Kollmeier et al. (2006), based on a sample of 407 AGNs from the AEGIS
survey, with BH masses estimated through a combination of line widths
and continuum luminosities; the agreement is reasonably good, though
Eddington ratios tend to be higher than the data at high redshift.
This shows that the reason why the STD model fails to produce bright
quasars relies in the low accretion rates stimulated by star
formation, while the main effect of winds is to force massive BHs to
accrete at high rates.
A similar prediction was given in Kauffmann \& Haehnelt (2000),
based on a parametric description of the accretion timescale.  Our
more refined modeling allows us to propose that this behaviour can be
obtained only by inserting quasar-triggered galaxy winds.
Finally, the predicted
dependence of the Eddington ratios on redshift and bolometric
luminosity can be useful to relate the accretion history of BHs to the
observed mass function of remnant BHs at $z=0$.


\section{Discussion}
\label{section:discussion}

The main message of this paper is that it is possible to include
accreting BHs in a complete galaxy formation model like {\gal} so as
to reproduce the main constraints on the AGN population from the
optical to the hard X-ray.  However, this good agreement is reached at
the cost of meddling with the very uncertain physics of
quasar-triggered galactic winds.  By presenting two different good
solutions, both based on motivated and plausible choices on how to
insert quasar winds, we intend to stress that the parameter space is
so wide, even in the restricted version presented here, that it is not
possible to constrain in a unique way the complex processes at play
simply by reproducing the statistical properties of the AGN
population.

\begin{figure}
\centerline{
\includegraphics[width=9cm]{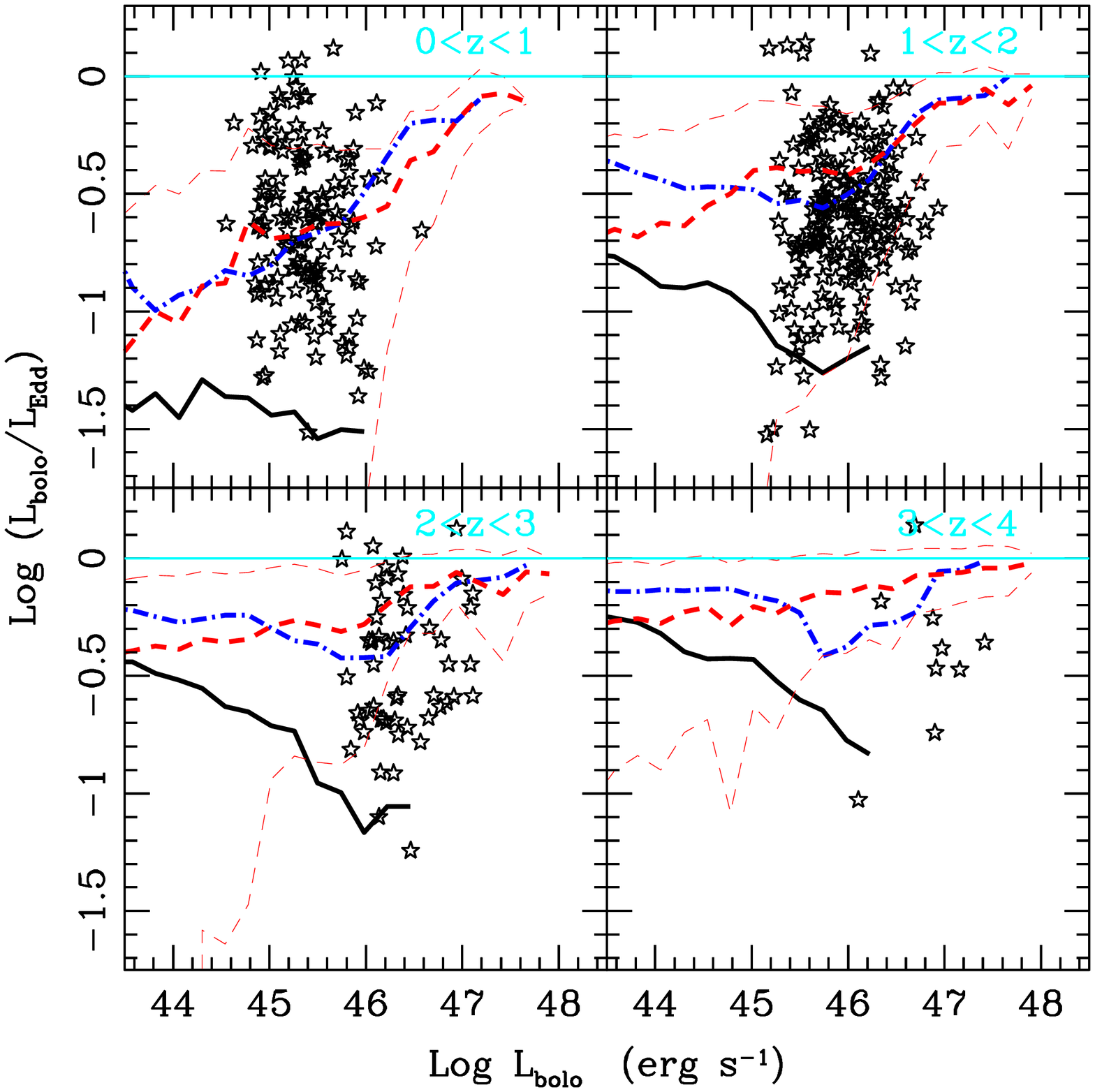}
}
\caption{Predicted average accretion rates in units of Eddington; 
thick lines refer to models as in figure~\ref{fig:hx_lf}. Thin
    lines represent the variance of the DW model, which is
    representative also for the other models. Stars refer to data from
    Kollmeier et al. (2006).
}
\label{fig:ed_lum}
\end{figure}

The second message of the paper is that, notwithstanding their
non-uniqueness, the only acceptable solutions we find are based on
quasar winds.  As clear in figure~\ref{fig:ed_lum}, the role of quasar
winds is to force a high accretion rate for the massive BHs and
reproduce the trend of increasing Eddington ratio with bolometric
luminosity, in better agreement with data.  Given the uncertainties
on the physics of these events, it is possible to use this result as a
suggestive clue on the role of winds but not as a strong evidence of
their necessity.  This caution is confirmed by the fact that most
papers cited in the Introduction are able to reproduce bright quasars
without advocating any such mechanism.  However, there is a deep
difference between our model, where the accumulation of low-$J$ gas is
connected to star formation, and most other models where some cold gas
is produced at a triggering event (mergers if not flybies) and
accreted at an arbitrarily specified fraction of the Eddington
rate.  In this regard our model can be directly compared only to that
of Granato et al. (2004), who proposed a very similar modeling of the
accretion onto the BH, but used a lower value of their parameter
equivalent to $f_{\rm lowJ}$; in their case the wind takes place when
the star-formation episode has consumed nearly all the gas, and their
winds do not limit strongly the BH mass. Their model should then be
more similar to our unsuccessful STD model.  But the real difference
between their approach and {\gal} relies in the treatment of star
formation and feedback; in particular,
they assume that the formation of an elliptical takes place in a
single episode and that the cooling gas goes to the bulge, while in
our model and in the same context the cooling gas would settle on a
disc\footnote{
In this context our choice of letting the cooled gas flow in the bulge
is not of particular relevance, as it is important only when most of
the galactic baryonic mass is in a bulge, so after the galaxy has
formed.}
until it loses angular momentum through bar instabilities, mergers or
feedback (see the discussion on the role of $\Sigma_{\rm lim}$ in
section~\ref{section:stellarfb}).  As a consequence, in {\gal} many of
the stars that end up in bulges are formed in discs and do not
contribute to the loss of angular momentum of gas (this is the main
reason for the scatter in the BH -- bulge relation).  Our higher
values of $f_{\rm lowJ}$ allow to have stronger accretion events and
higher BH masses at high redshift, but then a limiting mechanism, like
quasar winds, is necessary to avoid to overshoot heavily the BH --
bulge relation.  In this regard, the modest overestimate of large BH
masses by the DW and AW\footnote{
In the AW case, where the discrepancy is more pronounced, we have
verified that this overestimate is due to many minor accretion events
that take place at very low Eddington rates and thus do not contribute
to the AGN activity.
} models could be fixed by a more careful study of the parameter
space, but we find hints that winds are a key ingredient to
achieve agreement with many observables.

Our model is also comparable to that of Cattaneo et al. (2005), who
relate the accretion rate to the star-formation rate in a similar way
as equation~\ref{eq:lowJ}, with their equivalent of the $\alpha_{\rm
lowJ}$ parameter equal to 1.5.  Unfortunately, their relatively small
box (100 Mpc$/h$ with $256^3$ particles) does not allow to reach
strong conclusion on the prediction of their model for bright quasars.

The third message is that kinetic feedback in star-forming bulges is a
very good candidate for the main mechanism responsible for the
downsizing of AGNs.  This point however needs some discussion.  An
overall good agreement between models (DW or AW) and data is obtained
only for $M_{\rm B}\ga 10^{11}$ {\msun} and $M_{\rm BH}\ga10^8$
{\msun}; at smaller values the BH -- bulge relation steepens
considerably, especially in the DW model.  We think that this
discrepancy points to an intrinsic problem of {\gal} that is shared
with most other galaxy formation models.  As shown in paper I, the
predicted stellar mass function of bulge-dominated galaxies does not
have a broad peak at $\sim10^{10}$ \msun, as suggested by
observations, but presents a power-law tail of small objects which is
just below that of discs; this problem is shared, for instance, by the
model of Croton et al. (2006).  Recently, Fontana et al. (2006), based
on the GOODS-MUSIC sample (Grazian et al. 2006), reconstructed the
stellar mass function of galaxies up to $z\sim4$ and compared their
results with N-body and semi-analytic models including {\gal}.  All
the models were found to overpredict the number density of small
galaxies at $z\sim1$; many of these galaxies are the likely
progenitors of small bulges.  Similar results are obtained with the
Garching model (De Lucia, private communication).  The mechanism of
kinetic feedback can suppress effectively star formation in small
bulges, but, as mentioned above, cannot hamper stars that form in
discs to get into bulges.  As a consequence, the formation of BHs is
suppressed but the number of bulges is not, and the BH -- bulge
relation steepens (remarkably, the steepening is slightly less evident
in the AW case, where the discs with high gas surface density are
transformed into bulges).  In other words, a good fit of the number of
AGNs and an overestimate of the number of bulges can only be
compatible with a steep BH -- bulge relation.

\begin{figure}
\centerline{
\includegraphics[width=9cm]{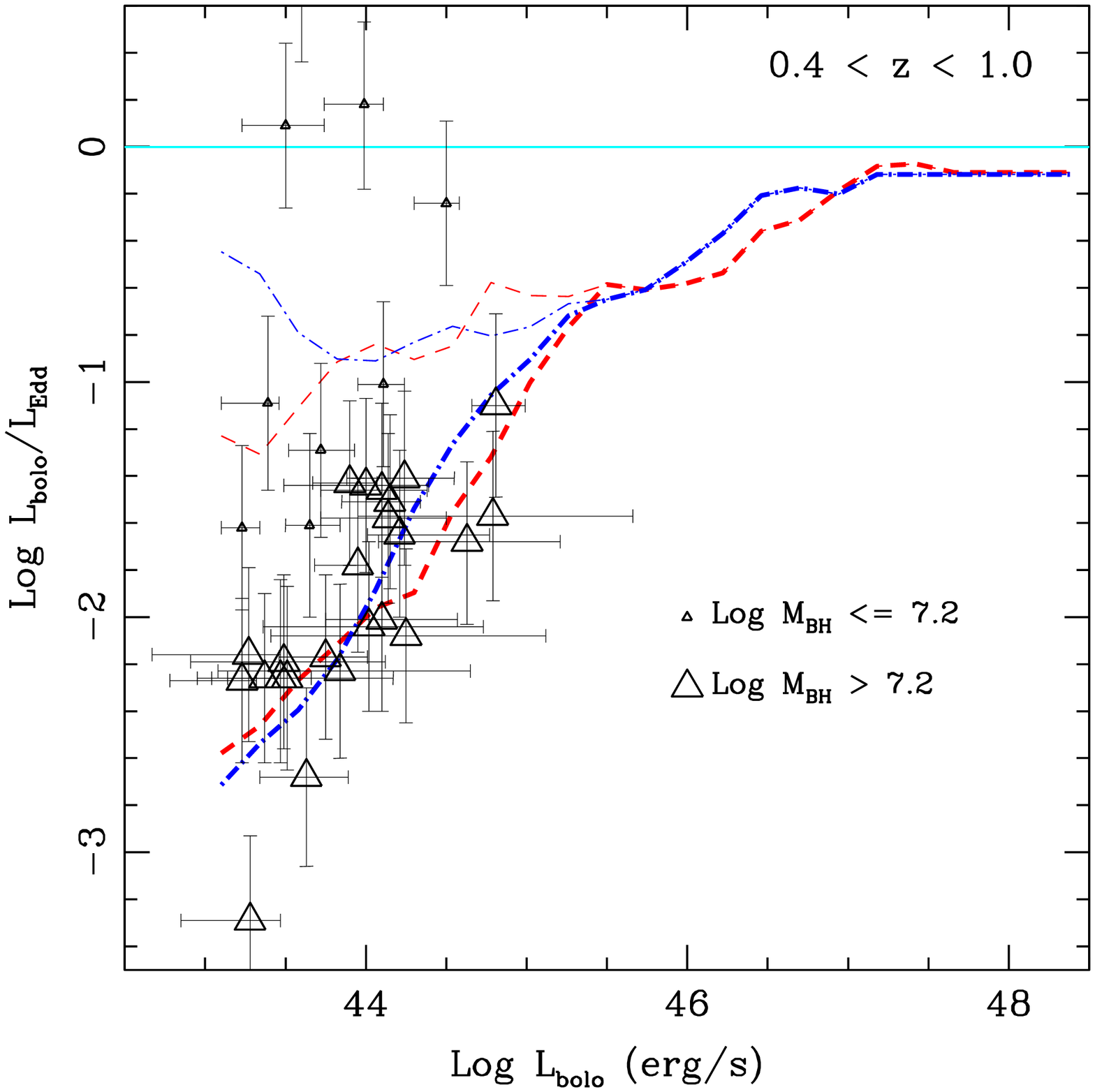}
}
\caption{Predicted Eddington ratios as a function of the bolometric
  luminosity for BHs more massive than $10^{7.2}$ {\msun} (thick
  lines, red dashed for the DW model, blue dot-dashed for the AW
  model) and for all BHs (thin lines, styles as above).  Data are from
  Ballo et al. (2006), large points refer to BHs more massive than
  $10^{7.2}$ {\msun}, smaller points to smaller BHs. The thick lines
  should be compared with the large points, the thin lines with {\em
  all} the points.}
\label{fig:ballo}
\end{figure}

The excess of small bulges is slightly over-corrected by kinetic
feedback, in that the resulting BH mass function is low at small
masses (figure~\ref{fig:mf_bh}).  This over-correction can be
explained as follows: kinetic feedback, which is applied at a rather
high level, limits the accretion on small BHs ($M_{\rm
BH}\simeq10^6-10^7$ \msun), but these BHs are anyway hosted in larger
bulges (figures~\ref{fig:bhbulz0} and \ref{fig:peng}), so they are
likely to have access to larger amounts of gas; then, a given (small
mass) BH can give rise to more (low-luminosity) AGN activity than
desired.  The constraint on the total accretion leads then to a
further underestimate of BH masses.  But if the small BHs are to
provide the correct amount of AGN activity, they should accrete on
average at a higher rate than observed.  To test this idea we compare
our predictions with the observations of Ballo et al. (2006), who
estimated stellar masses, bolometric luminosities and BH masses (based
on the BH -- bulge relation at $z=0$) for a hard X-ray-selected sample
of faint AGNs hosted in galaxies at $0.4<z<1$.  Figure~\ref{fig:ballo}
shows the Eddington fraction of the DW and AW models compared to the
Ballo et al. (2006) data for $M_{\rm BH}>1.6\times10^7$ {\msun} and
for all BH masses.  In particular, the thick lines show model
predictions with the same mass limit; these are in very good agreement
with the corresponding observational points (the larger triangles).
This means that the accretion rate of these BHs is correctly
reproduced by the model.  This result does not depend on the details
of the quenching of cooling flows by AGN feedback, as the involved BH
masses are rather small; indeed, identical results are obtained in the
higher resolution box with the standard quenching procedure.
Besides, the accretion rate of all BHs (thin lines), which is
dominated by the smaller objects, is tendentially larger than observed
(the thin lines must be compared with both large and small triangles).
This trend is not very strong, and could be due to a lack of small BHs
in the dataset; however, as discussed in the Ballo et al. (2006)
paper, the hard X-ray selection makes this possible bias weak if not
absent.  This comparison then confirms that, to compensate for the
excess of small bulges, our model generates too few small BHs
($M\la10^7$ \msun) that, being hosted in relatively larger bulges,
accrete at a higher rate.  The kinetic feedback mechanism remains then
a very good candidate for downsizing AGNs, but the value proposed here
of 60 {\kms} is probably overestimated.

The discrepancy described above points to some missing feedback
mechanism active at high redshift able to downsize the small
star-forming discs in the same way as kinetic feedback does for the
small bulges.  Bulge formation by feedback, explained in
section~\ref{section:stellarfb}, helps but does not solve the problem,
so other mechanisms are required.

Other authors have obtained good agreement between the prediction of
similar semi-analytic models and data in terms of the downsizing of
the AGN population, so it is worth wondering if other mechanisms than
kinetic feedback can give similarly good results.  In the case of
Granato et al. (2001, 2004), the downsizing is obtained by delaying
the shining of quasars, as originally suggested by Monaco et
al. (2000).  This delay is justified by stellar feedback, but the
model does not make a distinction between thermal and kinetic
feedback.  Menci et al. (2003, 2004, 2006; see also Vittorini et
al. 2006) obtain similarly good results, although the level of
predicted downsizing may be slightly less than that required by data
(Menci, private communication). In their case the downsizing is
obtained by stellar feedback (parameterized as usual with a $\beta$
coefficient which is a power-law of the disc velocity) and by the
insertion of galaxy-galaxy interactions as a further trigger of AGN
activity, a mechanism which penalizes the small satellites.
Similarly, Cattaneo et al. (2005) with a standard feedback recipe and
an accretion rate similar to equation~\ref{eq:lowJ} (but without
reservoir) obtain a roughly good match to the optical LF of quasars.
In all these cases stellar feedback is treated in a simple and
parametric way.  As shown and discussed in Monaco (2004a) and Paper I,
feedback in discs can lead to the reheating of an amount of cold gas
roughly equal to the star-formation rate, which means $\beta\simeq 1$,
so the only way to have more cold gas ejected to the halo is to {\em
accelerate} it more than {\em heating} it.  We are then rough
agreement with the papers mentioned above in stating that the
suppression of faint AGNs at high redshift is due to stellar feedback,
but our more refined modeling allows us to draw stronger conclusions
on the details of the physical mechanism at play.


\section{Conclusions}
\label{section:conclusions}

This paper belongs to a series devoted to presenting {\gal}, a new
model for the joint formation and evolution of galaxies and AGNs, and
is focused on the role of feedback in shaping the observed properties
of AGNs and their relation with galaxies.  Our analysis confirms that
models based on the $\Lambda$CDM cosmogony are able to roughly
reproduce at the same time both the properties of AGNs (presented
here) and the properties of galaxy populations (presented in paper I,
Fontana et al. 2006 and Fontanot et al. 2006b).

This model has been extended, through a careful modeling of AGN SEDs,
to predict many QSO properties, especially in the X-rays.  It
reproduces nicely LFs and number counts in the soft and hard X-ray
bands, and the measured background from 0.5 to 300 keV, with a
possible underestimate of the peak which confirms the claim for a
significant though not dominant population of Compton-thick sources.
This implies that the bulk of AGN accretion is roughly reproduced.  

The agreement between model and data allows us to draw these
conclusions.

{\em Quasar triggered winds are necessary in our model to reproduce
the number density of bright quasars.}  The parameter space of quasar
winds is discouragingly wide; within a very limited sub-space,
characterized by ``only'' six parameters, we find two good solutions,
which is a clear sign that our knowledge on the physics of accretion
onto BHs and their interaction with galaxies is still too poor to draw
firm conclusions.  In any case, the idea that quasar winds are a
necessary ingredient of galaxy formation is worth pursuing.  The two
wind solutions that we propose are based on the same triggering
criterion for the quasar wind, requiring that (i) the accretion rate
is high enough to perturb the ISM, (ii) the gas mass to remove is not
too high, (iii) the AGN is accreting in a radiatively efficient mode.
The first solution is based on a tilt of the relation between bulge
star-formation rate and loss rate of angular momentum, compensated by
``drying winds'', where the kinetic energy injected by the central
engine causes a complete removal of ISM from the host bulge.  The
second solution is based on the ``accreting wind'' mechanism proposed
by Monaco \& Fontanot (2005), where the wind is generated throughout
the ISM, so that a fraction of the cold gas is compressed to the
center and stimulates further accretion.  A good fit of the AGN LFs is
then obtained by assuming, as explained in paper I, that discs with
high gas surface density lose angular momentum (and then become
bulges) because of the expected change in the feedback regime (Monaco
2004a) and the consequent increase of the velocity dispersion of
clouds (as observed by F\"orster Schreiber et al. 2006).

{\em We propose that kinetic feedback in star-forming bulges is a very
good candidate for downsizing the AGN population.}  Indeed, the high
velocity dispersion of cold gas in star-forming bulges can lead to a
massive removal of ISM, leading to a suppression of less luminous AGNs
at high redshift, while larger elliptical galaxies can retain their
gas more easily.  At lower redshift larger BHs may accrete at lower
rates, giving rise to the low-luminosity population dominating the
X-ray background.

{\em We make the following predictions:} (i) the BH-bulge relation is
already in place at high redshift, though high BH masses in relatively
small bulges may be found at $z>2$ if quasar winds are active; (ii)
the Eddington ratios of accreting BHs depend both on redshift and
bolometric luminosity, with low values for the faint AGNs at $0<z<1$
that give the bulk of the hard X-ray background (consistent with Ballo
et al. 2006); (iii) most BH mass is acquired through accretion more
than through mergers, especially if quasar-triggered winds are
switched on.

{\em There is an important point of discrepancy between our model and
data, in that the predicted BH -- bulge relation at $M_{\rm
B}<10^{11}$ \msun, or $M_{\rm bh}<10^8$ \msun, is significantly
steeper than observed;} in the same mass range the predicted mass
function of BHs is too low.  To our understanding, this is connected
to a known excess of small bulges with respect to observations (see
paper I) and to the excess of small-mass galaxies predicted by most
(N-body or semi-analytic) models at $z\sim1$ (Fontana et al. 2006).
We have tested this idea by comparing our model predictions of the
Eddington ratios of faint AGNs at $0.4<z<1$ with the data of Ballo et
al. (2006); while the agreement is excellent for $M>1.6\times10^7$
\msun, smaller BHs accrete at a higher rate than observed, and this
compensates for the lack of smaller BHs.  This discrepancy highlights
that while kinetic feedback is efficient in downsizing AGNs by
quenching the star formation in bulges, stars in small ellipticals may
be born in discs, so that another mechanism is needed to produce the
required level of downsizing of bulges.

In conclusion, in developing this work we have gained interesting
insight into the complex problem of the cosmological rise of AGNs,
highlighting two promising astrophysical mechanism (kinetic feedback
in star-forming bulges and quasar-triggered galaxy winds) and the need
for a third, unknown one to improve the downsizing of elliptical
galaxies.  However, our poor understanding of the underlying physics
hampers more robust conclusions.  This demonstrates once again that
the field of joint galaxy and AGN formation is observationally-driven
and we need strategies to single out the physical mechanisms at play.
In particular, the main degrees of freedom in the theory are related
to the connection between BH accretion and star formation, which can
be constrained by estimating accretion rates, star-formation rates and
BH masses in active galaxies; to the quasar-triggered winds, which can
be constrained by observing warm and cold absorbers or Lyman-$\alpha$
blobs associated to quasars; and finally to the nature of stellar
feedback, which can be constrained by detailed observations of
starburst galaxies.  The next generation of telescopes will provide
suitable tools to assess these topics.

\section*{Acknowledgments}
We warmly thank Lucia Ballo for her permission to use her data prior
to publication and for many enlightening discussions.  We thank Andrea
Merloni for discussions.  PM thanks the Institute for Computational
Cosmology of Durham for hospitality.

{}

\bsp

\label{lastpage}

\end{document}